\documentclass[
 preprint,
 amsmath,amssymb,
 aps,
]{revtex4-2}
\usepackage{todonotes}
\usepackage{graphicx}
\usepackage{dcolumn}
\usepackage{bm}
\usepackage{subcaption}
\usepackage{tikz}
\usepackage{MnSymbol}

\begin{document}


\title{Learning subgrid interfacial area in two-phase flows with regime-dependent inductive biases}






\author{Anirban Bhattacharjee}
\email{abhattac315@gatech.edu} 
\affiliation{Flow Physics and Computational Science Lab, Georgia Institute of Technology, Atlanta, GA 30332, USA}

\author{Luis H. Hatashita}
\affiliation{Flow Physics and Computational Science Lab, Georgia Institute of Technology, Atlanta, GA 30332, USA}

\author{Suhas S. Jain}
\email{suhasjain@gatech.edu} 
\affiliation{Flow Physics and Computational Science Lab, Georgia Institute of Technology, Atlanta, GA 30332, USA}

\begin{abstract}
The reliability of machine learning in multiscale physical systems depends on how physical structure is embedded into the learning process. We investigate this in the context of turbulent multiphase flows, focusing on the prediction of subgrid interfacial area density, a key quantity governing interphase transport that remains unresolved in large-eddy simulations.
In this work, we develop and evaluate two machine learning subgrid closure models to predict the three-dimensional subgrid interfacial area density: a purely data-driven 3D encoder-decoder network, and a physics-constrained variant regularized by a fractal geometric prior.
Across a range of Weber numbers, the physics-based model improves predictive accuracy, reduces error variance, and suppresses nonphysical artifacts relative to purely data-driven approaches.
We also show that these gains are regime-dependent: the embedded inductive bias enhances generalization in corrugation-dominated regimes where its underlying assumptions hold, but becomes ineffective in fragmentation-dominated regimes characterized by topology change and droplet breakup.
These results reveal a broader principle for scientific machine learning: the utility of physics-informed models depends not only on the presence of inductive bias, but on its alignment with the governing physical regime. This suggests a path toward regime-aware learning frameworks for the modeling of complex multiscale systems.
\end{abstract}
\maketitle
\section{\label{sec:intro}Introduction}
Machine learning (ML) has emerged as a powerful tool for modeling complex physical systems, yet its reliability in multiscale regimes remains fundamentally limited by poor generalization outside the training distribution~\cite{choi:2025}. 
To explore the utility of physics-informed learning in such complex systems, this work uses two-phase flows as the testbed.
Such flows are observed in a wide range of settings, from wind gusts creating wave crests to atomizers generating small droplets~\cite{chan:2021,chigier:1979}. The distinct phases may or may not interact with each other, namely, exchange mass, momentum, and energy. For the former example, when the waves break, part of the air is engulfed by the water, generating small bubbles. Part of the oxygen and carbon dioxide contained therein is transferred to the large water body, both of which are essential for maintaining the bio-cycle of rivers and oceans~\cite{chan:2021}. Atomizers, used in printing, sprays in agriculture, and fuel injectors, create homogeneous distributions of droplets, yielding a more uniform absorption or evaporation of the liquid phase~\cite{chigier:1979}. Bubbly flows have been studied under microgravity environments due to their significance in the development of thermal management systems in spacecraft applications~\cite{takamasa:2003}. Similarly, high-speed multiphase flows play a critical role in many aerospace applications, including rocket combustion and hypersonic flight~\cite{stoffel:2023}. One important quantity in such flows is the interfacial area between the phases, which will dictate the total amount of mass, momentum, and energy transfer. 

Numerical simulations of such flows are beneficial for the efficient design and analysis of complex systems. In an industrial setting, where multiple parametric studies are required to optimize the final design, the state-of-the-art approach is via Reynolds-averaged Navier-Stokes (RANS) simulations, solving for the mean flow statistics with adequate closures. However, in a research setting, high-fidelity simulations, such as direct numerical simulation (DNS), are often preferred for their accuracy and limited use of models. Regarding computational cost, the latter is significantly more expensive, almost always requiring access to high-performance computing clusters. However, in terms of accuracy, RANS will not be able to reliably generalize its predictions when the models are off their respective ranges of application, whereas DNS does not rely on such models. With the current increase in computational power and availability of graphics processing units (GPUs), mid-fidelity simulations, such as large-eddy simulations (LES), provide a middle ground of cost and accuracy. In LES, the instantaneous flow is solved with closures to account for scales smaller than the computational grid (filter scale). For instance, an eddy viscosity-type model is typically used in single-phase turbulent flows to close the turbulent stresses arising from the nonlinearities of the Navier-Stokes equations in the LES framework~\cite{smagorinsky:1963}. In two-phase flows, on the other hand, subgrid closures often depend on the approach used to track/capture the interface and are less established. 

\begin{figure}
    \centering
    \begin{tikzpicture}
        \matrix[column sep=2pt] (m) {
            \node[inner sep=0] (A0) {\includegraphics[width=0.28\textwidth]{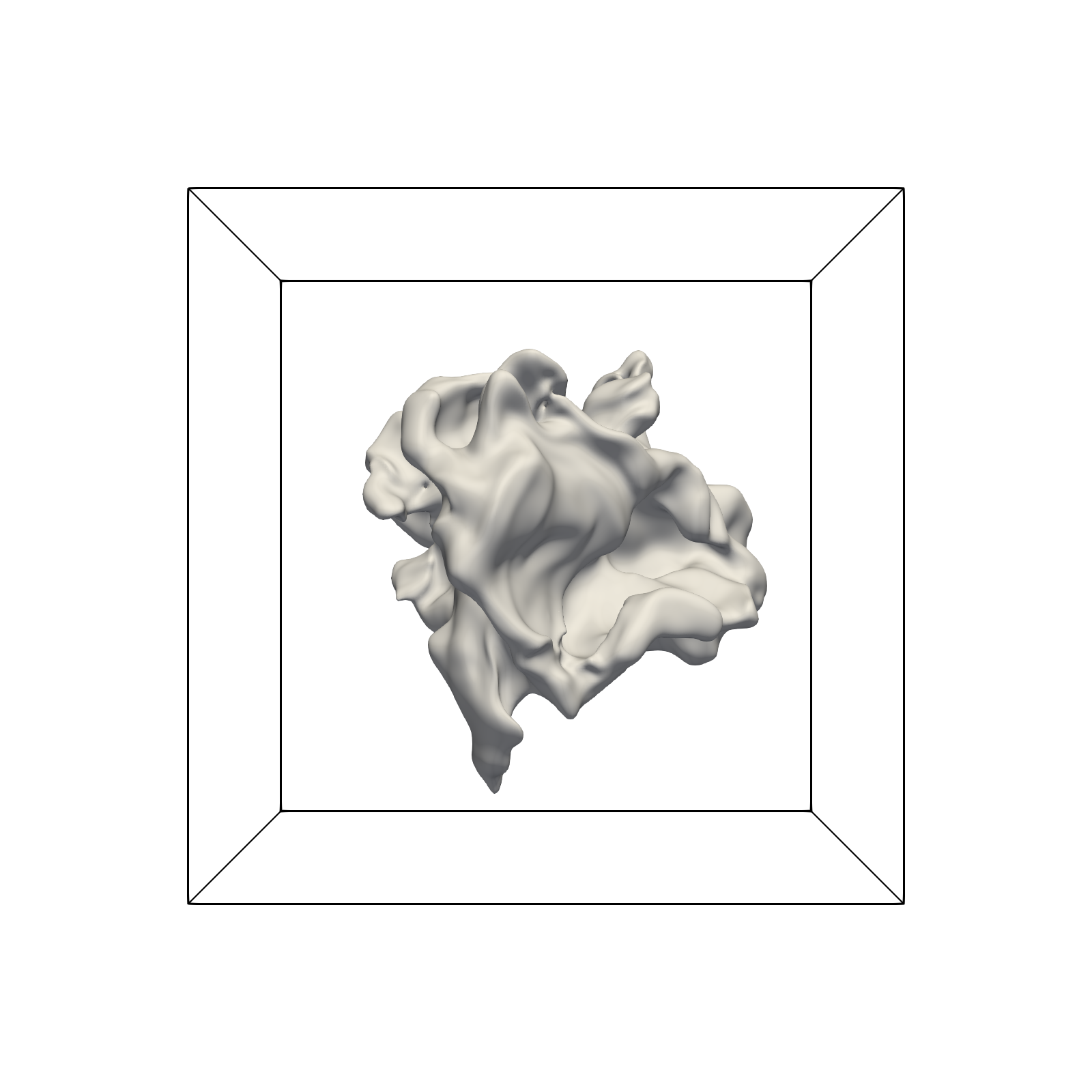}}; &
            \node[inner sep=0] (A1) {\includegraphics[width=0.28\textwidth]{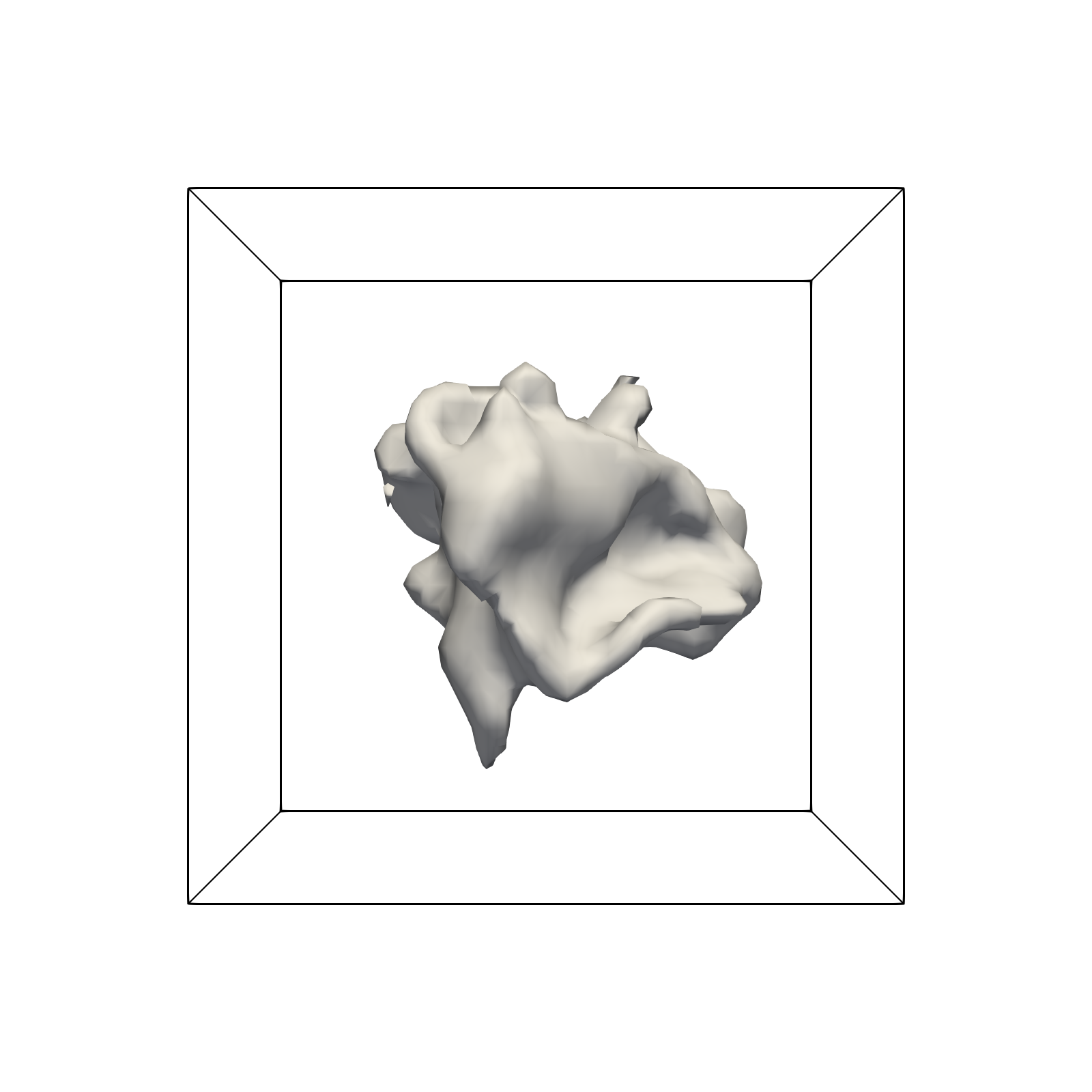}}; &
            \node[inner sep=0] (A2) {\includegraphics[width=0.38\textwidth]{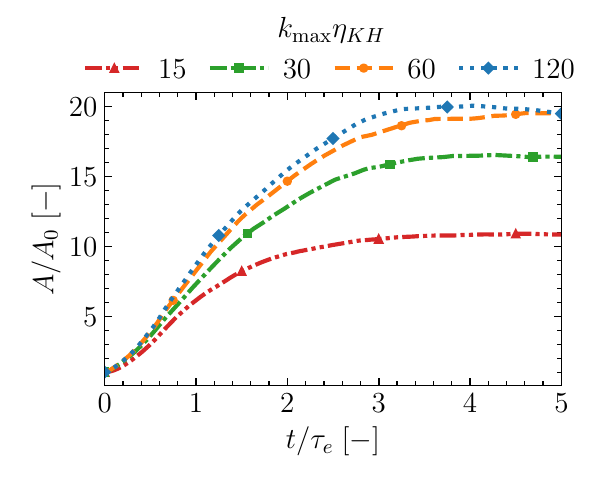}}; \\
        };

        \begin{scope}[shift={(m.south west)}, x={(m.south east)}, y={(m.north west)}]
            
            \node[black, font=\boldmath, anchor=south] at (0.03, 0.0) {(a)};
            2
            \node[black, font=\boldmath, anchor=south] at (0.32, 0.0) {(b)};
2
            \node[black, font=\boldmath, anchor=south] at (0.62, 0.0) {(c)};
            
        \end{scope}
    \end{tikzpicture}
    \caption{Illustration of a corrugated droplet in turbulence, from (a) a DNS simulation, and (b) an LES simulation on a $8\times$ coarser grid. The corrugations absent in the LES simulation result in lower interfacial area, which are to be modeled at the subgrid level. (c) The time evolution of total interfacial area on various grids.}
    \label{fig:DNSvLES}
\end{figure}
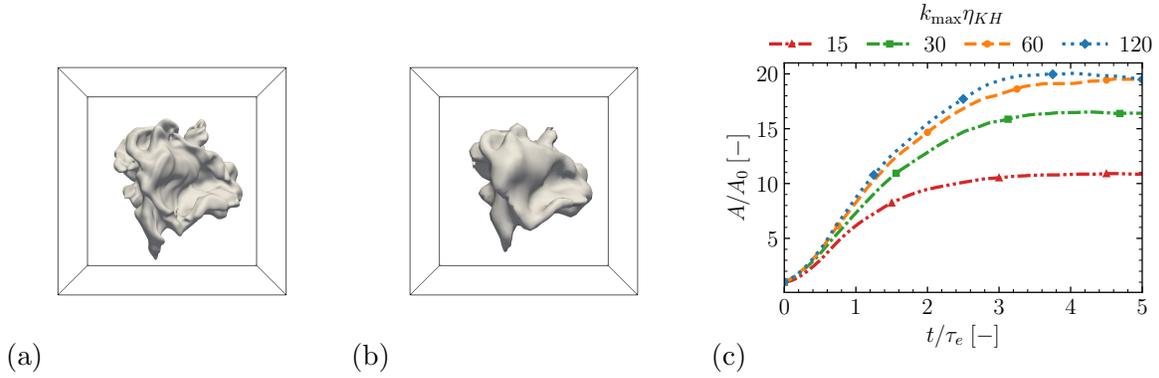

Interfacial area density is defined as the ratio of total interfacial area within a unit of volume. Thus, given a specific volume, \textit{e.g.}, the grid cell, one can always recover the total interfacial area by integrating the area density over the control volume. Total interfacial area thereby follows by integrating the interfacial area density over the entire domain, a key quantity for estimation of the total transfer of mass, momentum, and energy. Figure~\ref{fig:DNSvLES}(a) illustrates how all the scales of the interface are captured in a DNS simulation, whereas the simulation in Fig.~\ref{fig:DNSvLES}(b), which is $8 \times$ coarser, fails to capture some of those corrugations. 
Figure~\ref{fig:DNSvLES}(c) shows time evolution of interfacial area on various grid resolutions, that represents 
an LES simulation where both the turbulence scales and interfacial scales are not resolved ($k_{\max} \eta < 1.5$ and $k_{max}\eta_{KH} = 15$), an LES simulation with all turbulent scales but not all interfacial scales resolved ($k_{max}\eta = 1.5$ and $k_{max}\eta_{KH} = 30$), and the real DNS simulation with all turbulent and interfacial scales resolved ($k_{max}\eta > 1.5$ and $k_{max}\eta_{KH} = 60, 120$), where $k_{max}\eta_{KH}$ is the interfacial resolution parameter discussed in~\cite{hatashita2025scalings}. It is evident that the coarser grids ($k_{max}\eta_{KH}= 30, 60$) fail to capture the interfacial area over time, demonstrating the need for a closure model for area.  

Models for this quantity are available in the literature~\cite{vallet:2001,lebas:2009,chesnel:2011,granger:2024}. Initially developed for RANS, these models represent the interfacial area density through a transport equation of an Eulerian quantity, which is desirable for larger-scale simulations, eliminating the need for capturing all the scales of the interface.
In addition to convective and diffusive terms in the transport equation, it contains both breakup and coalescence models to account for the formation and destruction of bubbles/droplets, respectively. Breakup and coalescence are often based on empirical relations and will fail to yield accurate results outside of their range of application. Although, it is possible to use a transport equation, a simpler alternative is chosen, \textit{i.e.}, the direct prediction of subgrid-scale interfacial area density. The motivation is to avoid having to model all unclosed terms separately. One such model is our recently developed, fractal-based LES model for interfacial area density \cite{hatashita:2025}. While such analytical models provide robust scaling laws, they often simplify the non-linear interactions inherent in turbulent flows; machine learning models can provide an alternative framework to map these complex interactions directly from resolved variables.

Recent advancements in machine learning have significantly influenced the field of turbulence modeling, which is evident in the development of data-driven subgrid closures for single-phase LES. 
Yet its reliability in multiscale regimes remains fundamentally limited by poor generalization outside the training distribution. A central challenge is how to embed physical structure into learning models in a way that improves predictive capability without constraining the model beyond the validity of the underlying physics.
Classical closure models for LES, like the Smagorinsky model~\cite{smagorinsky:1963}, are based on the eddy-viscosity hypothesis which are stable but fail to capture important phenomena like backscatter and show a poor correlation for the subgrid stress term, whereas alternative models like the scale-similarity model~\cite{bardina:1980} predict backscatter and an improved correlation for the closure term, but may under-predict dissipation leading to numerical instability. To bridge this gap, early data-driven efforts employed empirical architectures like standard convolutional and fully-connected neural networks~\cite{beck:2019, wang:2018} to map resolved flow variables directly to subgrid stresses. These initial works, however, were agnostic to the underlying physics of the system. Therefore, interest in this field has shifted to developing physics-informed models which incorporate inductive biases such as the use of Tensor Basis Neural Network~\cite{wu:2025} which follows Galilean invariance, frame rotation equivariance, and locality, and accounts for appropriate backscatter and dissipation in the flow.

While the field of single-phase ML has seen the use of these physical constraints, the application of ML to three-dimensional multiphase turbulence remains sparse and mathematically unconstrained. Recent reviews show that there has been increasing interest in the use of machine learning for multiphase flows~\cite{basha:2024, cao:2026}. These reviews highlight a wide range of applications, such as developing reduced-order models to predict the evolution of slug flows~\cite{heaney:2022}, developing surrogate models for the design of flow reactors~\cite{savage:2024}, or utilizing fully connected neural networks for interface tracking~\cite{qi:2019, patel:2019, onder:2023}. 
Recently,~\cite{zuzio:2026} modeled the interfacial area from resolved volume fractions for the specific case of a liquid jet. 
While these ML advancements in multiphase flows focus on macroscopic flow evolution, they do not address the challenges of subgrid modeling for interfaces. Since these methods focus on bulk transport phenomena or features already fully resolved by the grid, the formulation of Eulerian subgrid closures for three-dimensional interfacial phenomena remains largely unexplored.
A key unresolved question is whether embedding physical constraints into machine learning models improves performance generally out of distribution, or whether their effectiveness depends on the validity of the underlying physical assumptions across regimes.

The objective of this work is to develop a physics-constrained machine learning model (and the corresponding data-driven model based on the encoder-decoder network with skip connections) for predicting the subgrid-scale total interfacial area and its local density in 3D turbulent multiphase flows; a task for which no prior work currently exists to the best of the authors' knowledge. We rely on actual DNS simulations that resolve all turbulent and interfacial scales, where the subgrid interfacial area is explicitly evaluated and modeled as a function of not just the volume fraction of one of the phases, but also the resolved turbulent scales in the flow.
We demonstrate how embedding a regime-dependent physical inductive bias, namely a fractal-based geometric prior, influences the learning and generalization of subgrid closures in turbulent multiphase flows. 
Here, “regime-dependent” refers to encoding how interfacial geometry varies with local Reynolds and Weber numbers, which control the distinction between corrugation-dominated and fragmentation-dominated regimes.

The structure of this paper is as follows: Section~\ref{sec:physics} characterizes the governing physics of 3D multiphase capillary interfaces in turbulence; Section~\ref{sec:formulation} describes the dataset, the machine learning model architecture, the fractal theory, characterizing the contribution of subgrid interfacial area, and introduces two different models, a data-driven one and another augmented with the fractal model bias; Section~\ref{sec:results} compares the results from both models; and Section~\ref{sec:conclusions} summarizes our work.

\section{Governing mechanisms and morphology of capillary interfaces in turbulence}
\label{sec:physics}

In the presence of turbulence, regimes of two-phase flows can be categorized into the inertia- and the viscous-dominated regimes~\cite{vankova:2007,ni:2024,hatashita2025scalings}. 
In addition to the capillary surface length scales, the range of scales of the carrier phase will delimit the breakup regimes, \textit{i.e.}, whether the smallest capillary surface length scale is smaller or greater than the smallest length scale of the carrier phase turbulence. 
The length scales of interest are: the smallest carrier phase turbulence scale, the Kolmogorov scale ($\eta$)~\cite{pope:2000}; the limiting capillary surface length scale if greater than $\eta$, the Kolmogorov-Hinze scale ($\eta_{KH}$); and the limiting capillary surface length scale if smaller than $\eta$, the Kolmogorov-viscous scale ($\eta_{KV}$).

The theory by Kolmogorov~\cite{kolmogorov:1949} and Hinze~\cite{hinze:1955} predicts a limiting scale in the inertial range of the carrier phase below which the capillary surface scales stop breaking ($\eta_{KH}$). In this regime, the balance is dictated by the dynamic pressure induced by turbulence fluctuations and the capillary energy. The theory predicts the limiting scale to be
\begin{equation}
\label{eq:def-eta-kh}
\eta_{KH} = \left (\frac{We_t^c \sigma}{\rho_c C_2 \epsilon^{2/3}} \right )^{3/5},
\end{equation}
where $We_t^c$ is the critical turbulent Weber number defining this limiting scale, $\sigma$ is the surface tension coefficient, $\rho_c$ is the carrier phase density, $C_2$ is the inertia range constant, and $\epsilon$ is the energy dissipation rate. The value of $We_t^c$ is a statistical representation of the demarcation of finite to zero breakup probability, although due to intermittency breakup can still occur below $\eta_{KH}$. $We_t^c$ is on the order of $1$, for instance, it has been reported as $\approx4.5$ in experiments~\cite{risso:1998} and $\approx3$ in numerical simulations~\cite{qian:2006}.
Based on the definition of $\eta_{KH}$ and the ratio of carrier phase turbulence scales $\eta/l$, one may construct the range of scales between the Kolmogorov-Hinze and the Kolmogorov scales (details in~\cite{hatashita2025scalings})
\begin{equation}
\label{eq:ratio-eta-kh-eta}
\frac{\eta_{KH}}{\eta} \sim \left ( \frac{We_t^c}{We_l} \right )^{3/5} Re_\lambda^{3/2},
\end{equation}
as a function of the large-scale Weber number ($We_l$) and the Taylor-microscale Reynolds number ($Re_\lambda$), to be defined in Section~\ref{subsec:data_description}. If $We_l$ is increased, the surface tension effects are weakened, allowing for smaller scales of carrier phase turbulence to break the capillary interfaces, thus decreasing the ratio $\eta_{KH}/\eta$. If $Re_\lambda$ is increased, the viscous effects are weakened, allowing for smaller scales of carrier turbulence to be developed ($\eta \searrow$), hence increasing the ratio $\eta_{KH}/\eta$.

The interaction of carrier-phase turbulence fluctuations with a capillary interface depends on the relative size of the interface compared with the Kolmogorov-Hinze scale. 
If surface tension is strong enough (for sub-Kolmogorov-Hinze scales, $d<\eta_{KH}$), the shape tends to a sphere, where the surface energy is minimized. Otherwise, for super-Kolmogorov-Hinze scales ($d>\eta_{KH}$), the surface tension is not sufficiently strong to overcome local deformations due to turbulence fluctuations, introducing corrugations and large deviations from spherical shapes. More recently, Cannon \textit{et al.} (2024)~\cite{cannon:2024} quantified curvature fluctuations across all scales, which characterize the changes in surface curvatures, representative of bumps and dimples, in addition to the natural curvature due to the surface tension equilibrium effect, tending to spherical shapes. Their results demonstrated that both sub- and super-Kolmogorov-Hinze scales present curvature fluctuations, providing further evidence of a fractal representation of the morphology of bubbles and droplets in turbulence.

The viscous-dominated breakup regime, in addition to the presence of a carrier phase turbulence range of scales, requires a sufficiently weak surface tension, such that the limiting interface scale is smaller than the Kolmogorov scale~\cite{shinnar:1961,vankova:2007}, thereby shifting the breakup mechanism from turbulence fluctuations to viscous shear stresses.
Achieving the viscous-dominated breakup regime is not trivial~\cite{ni:2024,hatashita2025scalings}, given that it either requires sufficiently high $\epsilon$~\cite{ni:2024} and an unattainable number of grid points for its numerical simulation~\cite{hatashita2025scalings}, therefore, limiting the current scope to the inertia-dominated regime.

\section{Formulation and datasets}
\label{sec:formulation}

\subsection{Dataset description}
\label{subsec:data_description} 

We have carried out direct numerical simulations (DNS) of forced two-phase isotropic turbulence~\cite{jain:2025} spanning two distinct regimes: a low Weber number regime at $Re_\lambda = 55$ with $We_l = 0.5 - 2.0$, and a high Weber number regime at $Re_\lambda = 87$ with $We_l = 4.0 - 14.0$. For both regimes, the combinations of $Re_\lambda$ and $We_l$ are chosen such that the turbulence resolution ($k_{\max}\eta$) and the inertia-driven breakup interface resolution ($k_{\max} \eta_{KH}$) are sufficient to capture the governing physics in two-phase turbulence, namely $k_{\max} \eta \geq 1.5$ and $k_{\max} \eta_{KH} \geq 60$, and to refer it to as DNS~\cite{hatashita2025scalings}. Here, $Re_\lambda$ is the Reynolds number based on the Taylor microscale ($\lambda$), and $We_l$ is the Weber number based  on the large energy-containing scale ($l$), each  respectively defined as
\begin{equation}
    Re_\lambda = \frac{u' \lambda}{\nu}, \hspace{1cm} We_{l} = \frac{\rho u'^2 l}{\sigma},
\end{equation}
where $u'$ is the root-mean-squared velocity, $\nu$ is the kinematic viscosity, $\rho$ is the density, and $\sigma$ is the surface tension coefficient. They represent the ratio of inertial forces to viscous forces, and the ratio of inertial forces to surface tension forces, respectively. The object of study is a pure capillary interface, represented by setting both density ($\rho_d/\rho_c$) and viscosity ($\mu_d/\mu_c$) dispersed-to-carrier ratios to $1$. The void fraction ($\langle \phi_d \rangle$, global volume of the dispersed phase over the domain volume) is set within the dilute regime with $\langle \phi_d \rangle\approx0.065$, by inserting a spherical dispersed phase with $D_0 = \pi$.

We are interested in evaluating the models across the two different regimes. In the low Weber number regime, the initial interface diameter ($D_0$) is smaller than the Kolmogorov-Hinze scale~\cite{kolmogorov:1949,hinze:1955} $(\eta_{KH})$, \textit{i.e.}, the surface tension forces are dominant, resisting breakup. In this regime, LES (and other lower-fidelity simulations) will fail to capture corrugations on the surface, as shown in Fig.~\ref{fig:DNSvLES}, due to insufficient interface resolutions $k_{\max}\eta_{KH}$. This results in the underprediction of the interfacial area, implying the need for its modeling at the subgrid level. 
Conversely, in the high Weber number regime, $D_0 > \eta_{KH}$, the inertial forces overcome surface tension for $D > \eta_{KH}$, causing further breakup/fragmentation. 
Table~\ref{tab:table1} highlights the Weber and Reynolds numbers for each regime, and the further splitting into training, validation, and test sets.

\begin{table}
\caption{\label{tab:table1}
$Re_\lambda$ and $We_l$ values for DNS data for training, validation, and test datasets both low and high $We_l$. 
}
\begin{ruledtabular}
\begin{tabular}{lccc}
\textrm{Case}&
\textrm{Dataset type}&
\multicolumn{1}{c}{$Re_\lambda$}&
\textrm{$We_l$}\\
\colrule
Low $We_l$ & Training & 55 & 0.5, 1.0, 1.25, 1.50, 1.75\\
 & Validation & 55 & 2.0\\
 & Test (Out-of-distribution) & 55 & 2.0\\
 & Test (In-distribution) & 55 & 1.25\\
 \hline
High $We_l$ & Training & 87 & 4, 6.5, 14.0\\
 & Validation & 87 & 9.5\\
 & Test & 87 & 9.5\\
\end{tabular}
\end{ruledtabular}
\end{table}

The flow consists of incompressible stationary homogeneous and isotropic turbulence (HIT) in a triply periodic $(2\pi)^3$ box, which is achieved by forcing the mixture turbulence kinetic energy to a constant via a linear forcing operator in physical space~\cite{jain:2025}.
A cubic domain with $128^3$ grid points ensures full resolution of all turbulent ($k_{\max}\eta=1.5$) and interface scales ($k_{\max}\eta_{KH}>60$) for $Re_\lambda = 55$ in the low $We_l$ regime ($<2.0$), while a finer $256^3$ grid is used to resolve the flow at $Re_\lambda = 87$ in the high $We_l$ regime, based on the grid-resolution estimates in~\cite{hatashita2025scalings}. Statistics are collected $5$ eddy turnover times after the insertion of the spherical dispersed phase, shown to be sufficient for the convergence of total interfacial area~\cite{hatashita2025scalings}. 
The dataset in each regime consists of $900$ flow snapshots where the training, validation, and test sets are split in a $10:1:1$ ratio. Further details of the numerical methods to gather the high-fidelity simulation data used in this work are provided in~\cite{jain2022accurate,jain:2025}, and of the choice of grid resolutions for resolved interfaces in turbulence in~\cite{hatashita2025scalings}.

\begin{figure}
    \centering
    \begin{tikzpicture}
        \matrix[column sep=2pt] (m) {
            \node[inner sep=0] (A0) {\includegraphics[width=\textwidth, trim={0 10em 0 0}, clip]{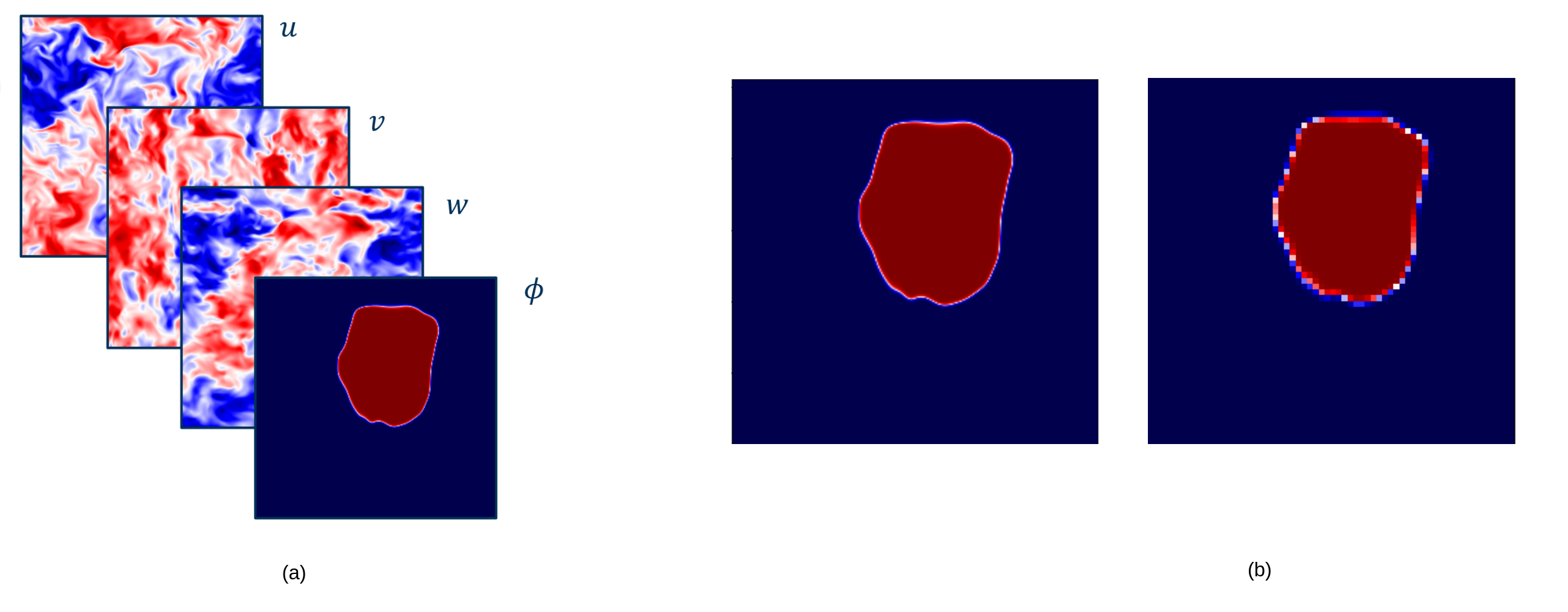}}; \\
        };

        \begin{scope}[shift={(m.south west)}, x={(m.south east)}, y={(m.north west)}]
            
            \node[black, font=\boldmath, anchor=south] at (0.0 , -0.1) {(a)};
            \node[black, font=\boldmath, anchor=south] at (0.45, -0.1) {(b)};
            
        \end{scope}
    \end{tikzpicture}
    \caption{(a) Dataset consists of velocity components and volume fraction obtained from DNS simulations, and (b) illustration of the effect of filtering and downsampling on the volume fraction field, showing the DNS field (left) and after filtering and downsampling by a factor of $4$ (right) in the low $We_l$ regime.}
    \label{fig:downsample} 
\end{figure}
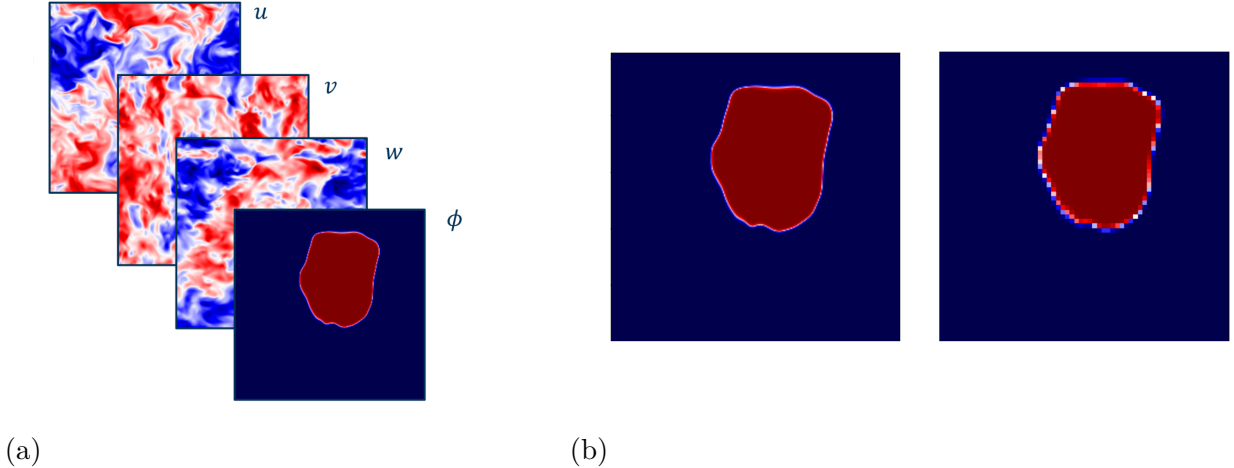

Each 3D snapshot has features $\{u, v, w, \phi\}$, where $u, v, w$ are the velocity components, and $\phi$ is the volume fraction of one of the phases at each cell as shown by their central planes in Fig.~\ref{fig:downsample}(a). The architecture aims to map resolved/LES features to a subgrid closure for interfacial area density, thus, the baseline features of the snapshots require further pre-processing, elaborated below. Namely,
each feature is filtered using a Gaussian filter and downsampled, both defined on a filter width of $\overline{\Delta}=4\Delta$ (where $\Delta$ is the DNS grid size) in the low $We_l$ regime, and on a filter width of $\overline{\Delta}=8\Delta$ in the high $We_l$ regime. 
For any feature $\mathbf{U}(\mathbf{x},t)$, the filtering operation is defined as a convolution of $\mathbf{U}$ with the filter kernel $\mathbf{G}$,
\begin{equation}
\label{eq:filt-def}
    \overline{\mathbf{U}}(\mathbf{x},t;\overline{\Delta}) = \int \mathbf{G}(\mathbf{r};\overline{\Delta}) \mathbf{U}(\mathbf{x}-\mathbf{r},t) d\mathbf{r},
\end{equation}
where the Gaussian filter kernel, based on the filter width $\overline{\Delta}$, is given by
\begin{equation}
\label{eq:gauss-filter-kernel}
    \mathbf{G}(\mathbf{r};\overline{\Delta}) = \sqrt{\frac{6}{\pi \overline\Delta^2}}e^{-\frac{6\mathbf{r}\cdot\mathbf{r}}{\overline\Delta^2}},
\end{equation}
such that $\overline{\mathbf{U}}$ is the filtered quantity. Downsampling is defined as an equidistant striding over the underlying LES computational space. While it can also be constructed as a projection over the LES grid, the former is chosen over the latter for its lower computational cost.
For instance, Fig.~\ref{fig:downsample}(b) shows a 2D slice of the procedure of filtering and downsampling of a baseline volume fraction field ($u,v,w$ are filtered and downsampled following the same procedure). Our machine learning model predicts the subgrid interfacial area density, which is given by
\begin{equation}
    \label{eq:area}
    \delta' = \langle {\left |\nabla \phi \right |}\rangle - \left | \overline{\nabla} \hspace{0.75mm} \overline{\phi} \right |,
\end{equation}
where $\left | \cdot \right |$ denotes the vector magnitude of a given quantity.
The DNS ``filtering and downsampling'' operations are constructed such that there is no loss of total interfacial area. Therefore, we propose the volume average operator $\langle \cdot \rangle$, defined as the ratio of the integral of the DNS interfacial area density $\delta$ over the LES cell by the LES cell volume, guaranteeing the preservation of total interfacial area. The resolved state follows from filtering and downsampling [Eq.~\eqref{eq:filt-def}] of the volume fraction field ($\overline{\phi}$), and construction of the resolved interfacial area density on the LES grid, \textit{i.e.}, the magnitude of the LES gradient operator ($\overline{\nabla}$) of the filtered and downsampled $\overline\phi$. The filter widths of the volume averaging ($\overline{\Delta}_{\rm DNS}$) and filtering/downsampling of the resolved LES field ($\overline{\Delta}_{\rm LES}$) operations are chosen to be the same.
Further details are presented in Appendix~\ref{apx:filtering}.

The inputs of the neural network are the filtered velocity fields, $\overline{u}, \overline{v}, \overline{w}$, and the filtered volume fraction field, $\overline{\phi}$, while the output is the subgrid interfacial area density field, $\delta'$, defined in Eq.~\eqref{eq:area}.
The proportional downsampling ratios (a factor of $4$ for the $128^3$ grid, and a factor of $8$ for the $256^3$ grid) are chosen to represent relevant LES grid point counts, whilst maintaining identical input dimensions, facilitating architecture construction.
Therefore, the inputs and output for our machine learning model across both regimes are of dimensions $32 \times 32 \times 32 \times 4$ and $32 \times 32 \times32 \times 1$, respectively, both of which are channel-wise normalized with a minimum-maximum scalar prior to training.

\subsection{Architecture}
\label{subsec:arch}

\begin{figure}
\centering
\includegraphics[width=1.2\textwidth]{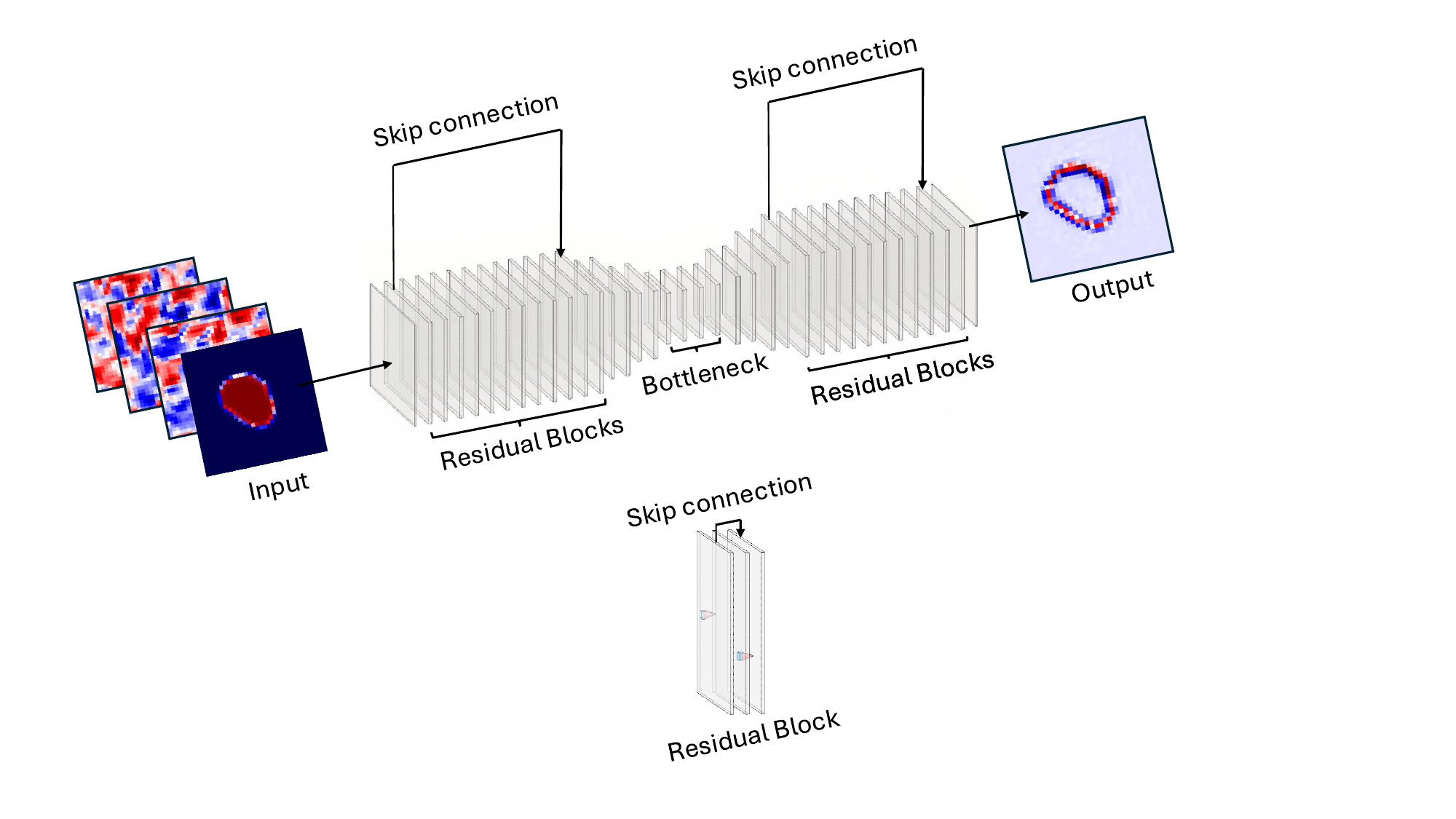}
\caption{\label{fig:Architecture} Autoencoder (encoder-decoder) architecture of our ML Model. The model contains residual blocks in both the encoder (left) and the decoder (right), and skip connections as shown inside and also across all the residual blocks. The input, output, and the layers have been shown as 2D images, but are 3D snapshots in our model.}
\end{figure}

We use a 3D autoencoder model with an encoder-decoder structure and skip connections based on the work of \cite{chung:2022} and \cite{glaws:2020}, which is known to perform effectively in predicting turbulence physics. The encoder carries a dimensionality reduction of the input into a latent space referred to as the bottleneck. It takes in an input with dimension $32 \times 32 \times 32 \times 4$, and first encodes it to a dimension of $4 \times 4 \times4 \times 64$. The number of channels is increased from $4$ to $64$ to preserve the information, and to compensate for the dimensionality reduction of the input. The encoding is done by multiple residual blocks, with each residual block consisting of two 3D convolutional layers, where only the first convolution operation is followed by a nonlinear ReLU activation as shown in Fig.~\ref{fig:Architecture}. Skip connections across each residual block help preserve the latent information from previous layers, and consequently the physics of different scales. 

Similarly, the decoder upsamples the $4 \times 4 \times4 \times 64$ bottleneck tensor to the $32 \times 32 \times 32\times 1$ output, which is the subgrid interfacial area density. The decoder also has residual blocks, with skip connections. We use a filter of size $3 \times 3 \times3$ for every operation, which resembles a stencil of neighboring points commonly used in algorithms for computational fluid dynamics. We use a filter stride of $1 \times 1 \times1$ or $2 \times 2 \times2$ depending on whether we want to maintain the dimension or downsample (or upsample) respectively. The autoencoder has 590,689 trainable parameters. We used Adam optimization~\cite{adam:2014} and $4000$ epochs to train the model.

\subsection{Data-driven and physics-based ML model}
\label{subsec:models}

The data-driven model maps the inputs to the subgrid interfacial area density which is the output with a nonlinear map $f_\theta$ given by
\begin{equation}
    \delta'_{\rm pred} = f_\theta(\overline{u}, \overline{v}, \overline{w}, \overline{\phi}),
\end{equation}
where $f_\theta$ is the encoder-decoder model used in this work and $\theta$ are the trainable parameters to be optimized. In the data-driven approach, the loss function, $\mathcal{L} (\theta)$, which is the mean square error between the predicted and actual subgrid area density obtained from DNS across all $N$ snapshots in the training dataset, is minimized and is given by
\begin{equation}
    \mathcal{L} (\theta) = \frac{1}{N}\sum_{i = 1}^N\left\|\delta'^{(i)}_{\rm pred} -  \delta'^{(i)}_{\rm true}\right\|^2.
\end{equation}
This model relies on the data alone, which can overfit and lead to poorer generalization on unseen data.

Thus, we further propose a physics-based model which uses the same encoder-decoder architecture, however, it also incorporates a regularization term that biases the parameters to satisfy the fractal theory described below. The loss function $\mathcal{L} (\theta)$ is now defined as
\begin{equation}
    \mathcal{L} (\theta) = \frac{1}{N}\sum_{i = 1}^N\left\|\delta'^{(i)}_{\rm pred} -  \delta'^{(i)}_{\rm true}\right\|^2 + \lambda_f \frac{1}{N}\sum_{i = 1}^{N}\mathcal{M}^{(i)}\left\|\delta'^{(i)}_{pred} -  \delta_f'^{(i)}\right\|^2,
    \label{eq:physics_loss}
\end{equation}
where $\delta'_{\rm true}$, $\delta_f'$, and $\delta'_{\rm pred}$ are the ground truth area, and the areas densities based on the fractal theory, and the machine learning model, respectively. The first term in Eq.~\eqref{eq:physics_loss} above is the data loss and the second term is the fractal theory based regularization term. $\lambda_f$ is a hyperparameter that indicates the weight of the physics term relative to the data term in the loss. We use a mask, $\mathcal{M}$, defined as
\begin{equation}
\mathcal{M}^{(i)} = 
\begin{cases} 
1 & \text{if } \delta'^{(i)}_{\rm true} > 0 \\
0 & \text{if } \delta'^{(i)}_{\rm true} < 0. 
\end{cases}
\end{equation}
The volume average and filtering/downsampling operations, described in Appendix~\ref{apx:filtering}, introduce both negative and positive subgrid areas around the interface due to the discrepancy in the interface scales, whereas the fractal model only predicts positive subgrid area due to corrugations. Therefore, we activate the mask and apply the physics regularizer only when the ground truth is positive (indicating physical subgrid areas).

Recently, we developed a physics-based subgrid model for the interfacial area using fractal theory \cite{hatashita:2025}. In that work, the interface is assumed to be a fractal, given its interaction with the carrier-phase turbulence, known to also be represented by a fractal dimension~\cite{sreenivasan:1986}. Nevertheless, due to the inherent scale separation around the Kolmogorov-Hinze scale ($\eta_{KH}$), there are different morphologies for sub- ($D<\eta_{KH}$) and super-Kolmogorov-Hinze ($D>\eta_{KH}$) scales, thus, we observed the existence of two distinct fractal dimensions, one for $D < \eta_{KH}$ (where the surface tension is more dominant, and the interface retains a nearly spherical shape), and the other for $D > \eta_{KH})$ (where the drop develops more corrugations, identified by curvature fluctuations). It was shown using this theory that the ratio of the total interfacial area $(A_\eta)$ to the resolved area $(A_{\overline{\Delta}})$ can be scaled as
\begin{equation}
    \frac{A_\eta}{A_{\overline{\Delta}}} \sim \left(We_{\overline{\Delta}}^{-3/5}Re_{\overline{\Delta}}^{3/4}\right)^{D_{<\eta_{KH}}- 2}\left(We_{\overline{\Delta}}^{3/5}\right)^{D_{>\eta_{KH}}- 2},
    \label{fractal1}
\end{equation}
where $We_{\overline{\Delta}}$ and $Re_{\overline{\Delta}}$ are the grid-based Reynolds and Weber numbers defined as
\begin{equation}
    Re_{\overline{\Delta}} = \frac{\overline{u}' \overline{\Delta}}{\nu}, \hspace{1cm} We_{\overline{\Delta}} = \frac{\rho \overline{u}'^2 \overline{\Delta}}{\sigma}.
\end{equation}

For the low Weber number regime, where the interface scale is smaller than the Kolmogorov-Hinze scale ($D,\overline{\Delta}<L<\eta_{KH}$, where $L$ is the domain length scale), Eq.~\eqref{fractal1} can be rewritten as
\begin{equation}
    \frac{A_\eta}{A_{\overline{\Delta}}} \sim \left(We_{\overline{\Delta}}^{-3/5}Re_{\overline{\Delta}}^{3/4}\right)^{D_{<\eta_{KH}}- 2}.
    \label{fractal2}
\end{equation}
The ratio of interfacial area can be shown to be equal to the ratio of interfacial area density,
\begin{equation}
    \frac{\langle \delta_\eta\rangle}{\delta_{\overline{\Delta}}} = \frac{A_\eta / \overline{\Delta}^3}{A_{\overline{\Delta}}/ \overline{\Delta}^3} \sim \left(We_{\overline{\Delta}}^{-3/5}Re_{\overline{\Delta}}^{3/4}\right)^{D_{<\eta_{KH}}- 2}.
    \label{fractal3}
\end{equation}
The difference between the ``filtered'' DNS interfacial area density ($\langle\delta_\eta\rangle$) and the resolved area density $(\delta_{\overline{\Delta}})$ yields the subgrid interfacial area density [Eq.~\eqref{eq:area}].
\begin{equation}
    \delta_f^{'} = \langle \delta_\eta\rangle - \delta_{\overline{\Delta}} \sim \left[\left(We_{\overline{\Delta}}^{-3/5}Re_{\overline{\Delta}}^{3/4}\right)^{D_{<\eta_{KH}}- 2} - 1\right]\delta_{\overline{\Delta}}.
    \label{fractal}
\end{equation}
As shown in \cite{hatashita:2025}, we take the dimension $D_{< \eta_{KH}} \approx 2.1$.
This formulation for $\delta_{f}^{'}$ is applied as a regularizer in the high $We_l$ regime as well, to test the regime-dependency of the inductive bias and because of the lack of closed-form geometric scaling laws in this regime. This allows us to evaluate whether a physical constraint continues to improve generalization in a regime, where its underlying assumptions are no longer strictly valid.
\section{Results}
\label{sec:results}

We evaluate the two models discussed above in the corrugation-dominated low $We_l$ regime, where the strong surface tension forces withstand breakup, and the high $We_l$ regime, with weaker surface tension forces and significant breakup. We also show the limitations of a standard $L_2$ regularizer, compared to the fractal-regularizer used in the physics-based model. 

\subsection{Low $We_l$ regime}
This section is focused on the low $We_l$ regime, where surface tension keeps the droplets largely intact. The datasets rely on snapshots filtered and downsampled from the DNS simulations at $Re_\lambda= 55$, with the training set covering five distinct Weber numbers $(We_l \in \{0.5,1.0,1.25,1.50,1.75\})$. To evaluate the generalization capabilities of the machine learning models, we studied two distinct test sets. First, to assess in-distribution performance, we created a set of entirely unseen snapshots at $We_l=1.25$, which represents a Weber number the models observed during training. Second, to test out-of-distribution extrapolation, we utilized a dataset at a higher, unseen Weber number $(We_l=2.0)$, dividing it into non-overlapping validation and test sets. Given the high computational cost of training $3D$ two-phase turbulence, hyperparameter tuning on the validation set was restricted to $\lambda_f$ in Eq.~\eqref{eq:physics_loss}. This parameter governs the primary trade-off between minimizing data-driven error and enforcing the physical constraint based on fractal theory. We used three different values ($\lambda_f \in \{0.001, 0.01, 0.1\}$), and selected the value of $0.1$ since it maximized $R^2$ on the validation dataset. While global loss convergence is tracked using the extrapolative $We_l = 2.0$ set, the final predictive performances of both models, including $R^2$ scores, were evaluated for both $We_l=1.25$ and $We_l=2.0$ sets.

\begin{figure}
\centering
\includegraphics[width=0.95\textwidth]{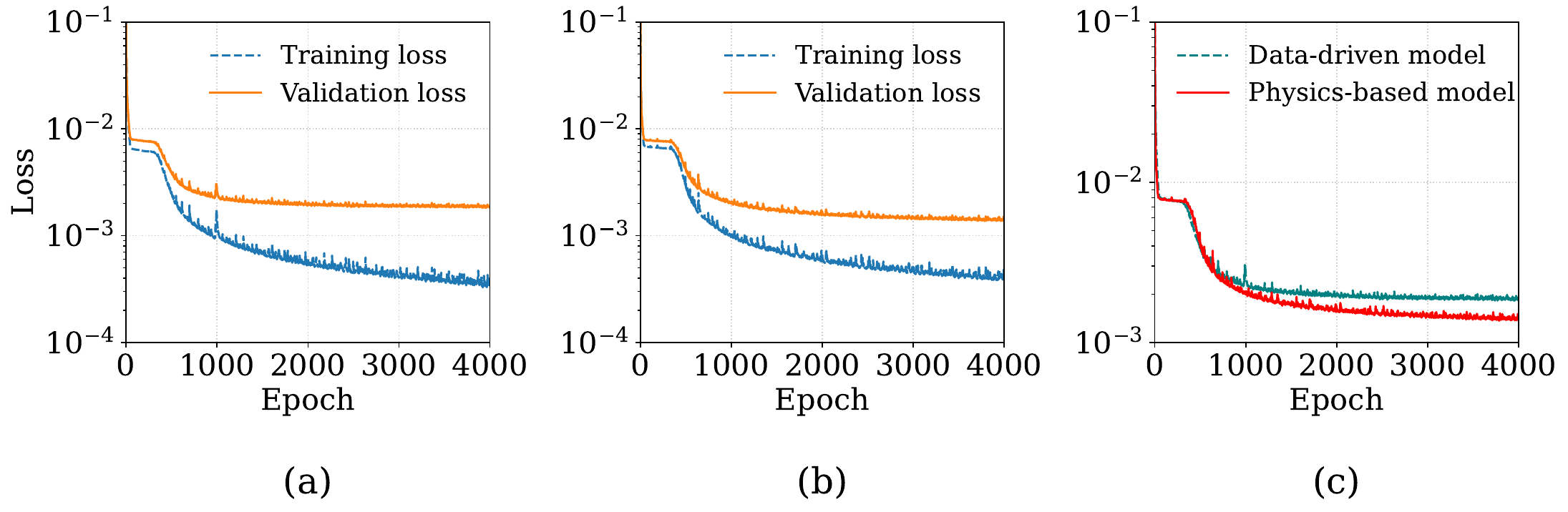}
\caption{\label{fig:Loss} Training and validation loss versus the number of epochs in the low $We_l$ regime for the (a) data-driven model and (b) physics-based model. In (c), the validation loss for both the physics-based model and the data-driven model are shown, and the physics-based model shows a better validation loss than the data-driven model across epochs, indicating better generalization.}
\end{figure}

Figure~\ref{fig:Loss} shows the training and validation losses. We ran both models for $4000$ epochs for a one-to-one comparison and to ensure the losses reached an asymptotic state. At the end of optimization, the data-driven model reached a training loss of $3.49\times10^{-4}$ and a validation loss of $1.87\times10^{-3}$ [Fig.~\ref{fig:Loss}(a)]. This order-of-magnitude discrepancy is expected for an extrapolation study, where the validation set resides at a higher Weber number $(We_l = 2.0)$ unseen during training. The final evaluation on the test set yielded a loss of $2.86\times10^{-3}$.

For the physics-based model, the final training and validation losses were $4.21\times10^{-4}$ and $1.42\times10^{-3}$ [Fig.~\ref{fig:Loss}(b)]. The training loss is marginally higher than that of the data-driven model due to the fractal-based regularization term in Eq.~\eqref{eq:physics_loss} which restricts the model from overfitting the training data to maintain physical consistency. This trade-off improves generalization as shown by a lower validation loss and subsequently a reduced test loss of $2.18\times10^{-3}$. Figure~\ref{fig:Loss}(c) shows that the inclusion of fractal regularization achieves a consistent improvement in generalization performance.

\begin{figure}
    \centering

    \includegraphics[width=\textwidth]{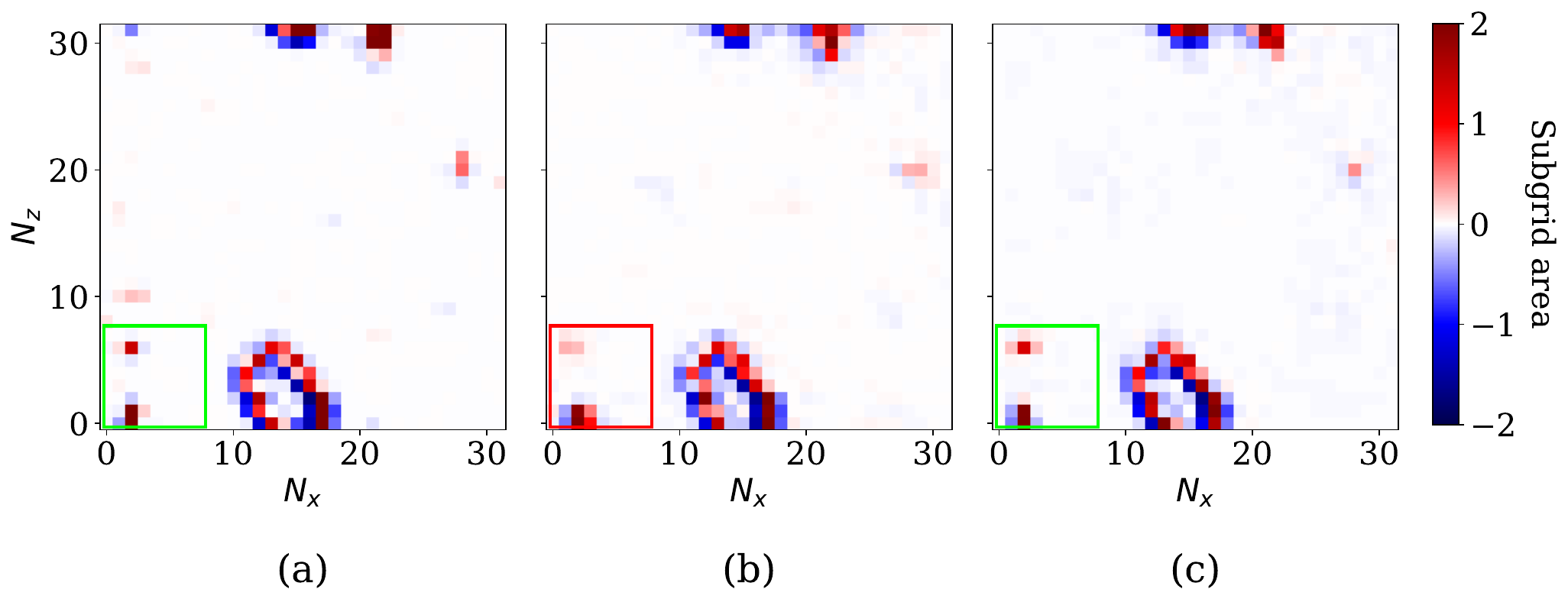}
    
    \vspace{2mm} 
    
    \includegraphics[width=\textwidth]{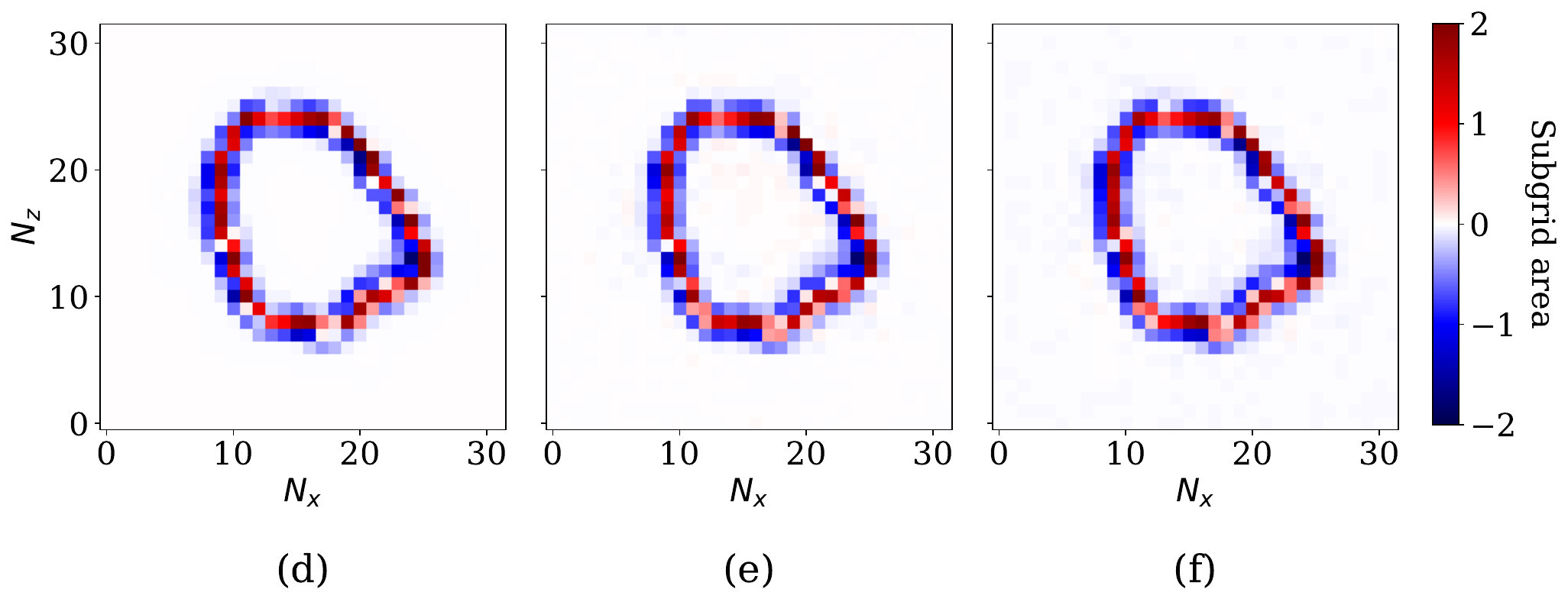}
    
    \caption{\label{fig:pred}Qualitative comparison between the prediction of subgrid interfacial area density from (a,d) the ground truth, (b,e) the data-driven model, and (c,f) the physics-based model. The top row shows $We_l = 2.0$ regime and the bottom row shows $We_l = 1.25$ regime. The central $x-z$ plane is shown in all cases.}
\end{figure}
Figure~\ref{fig:pred} shows the predicted local subgrid interfacial area density from both models against the ground truth for representative cross-sectional planes of one snapshot each sampled from the out-of-distribution and the in-distribution sets. 
While both models successfully capture the large-scale interfacial structures at $We_l = 1.25$ (bottom row), their predictions diverge at the higher Weber number set ($We_l = 2$).

\begin{figure}
    \centering
    \begin{subfigure}[b]{0.42\textwidth}
        \centering
        \includegraphics[width=\textwidth]{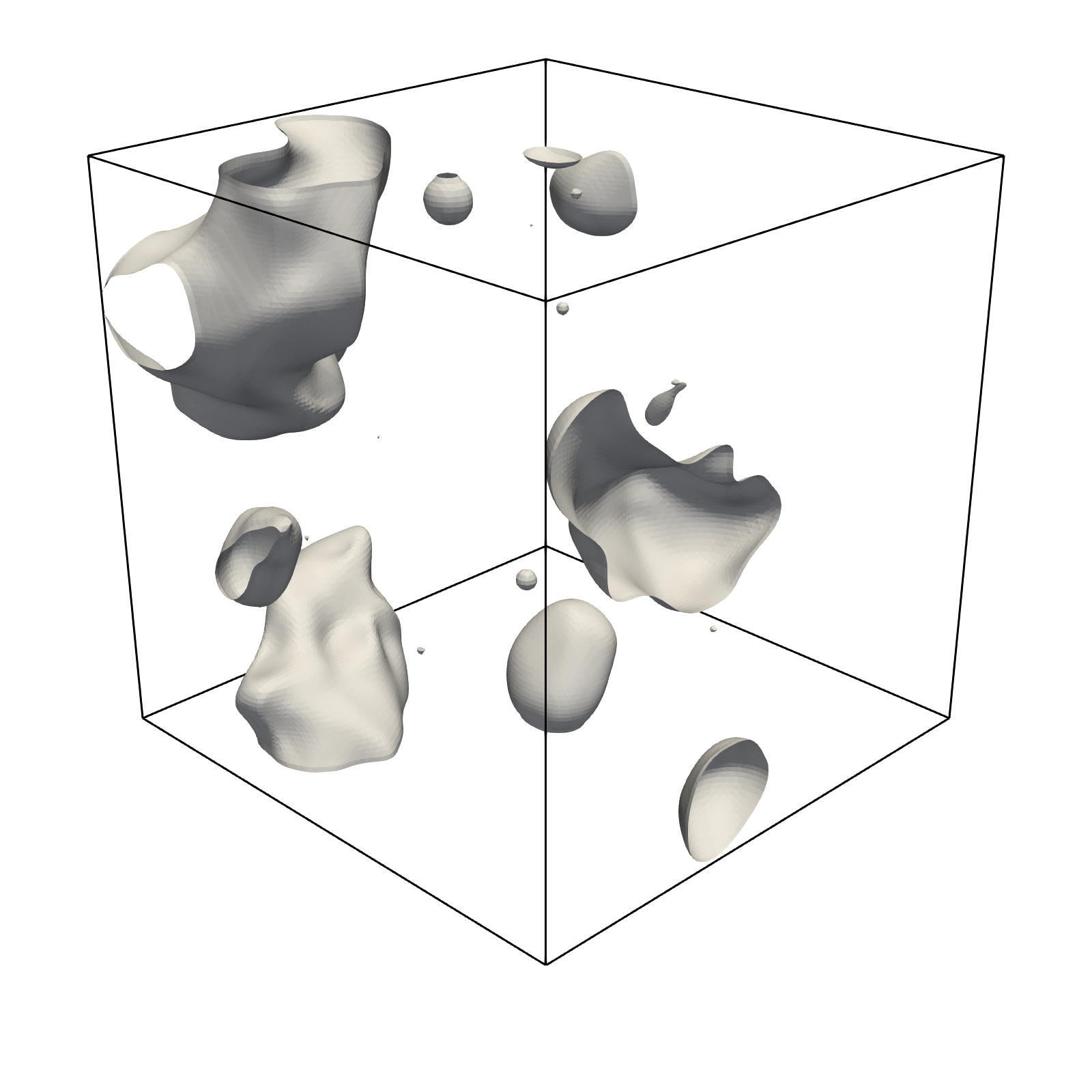}
        \caption{Low $We_l$ regime}
        \label{fig:phi 2_0}
    \end{subfigure}
    \begin{subfigure}[b]{0.42\textwidth}
        \centering
        \includegraphics[width=\textwidth]{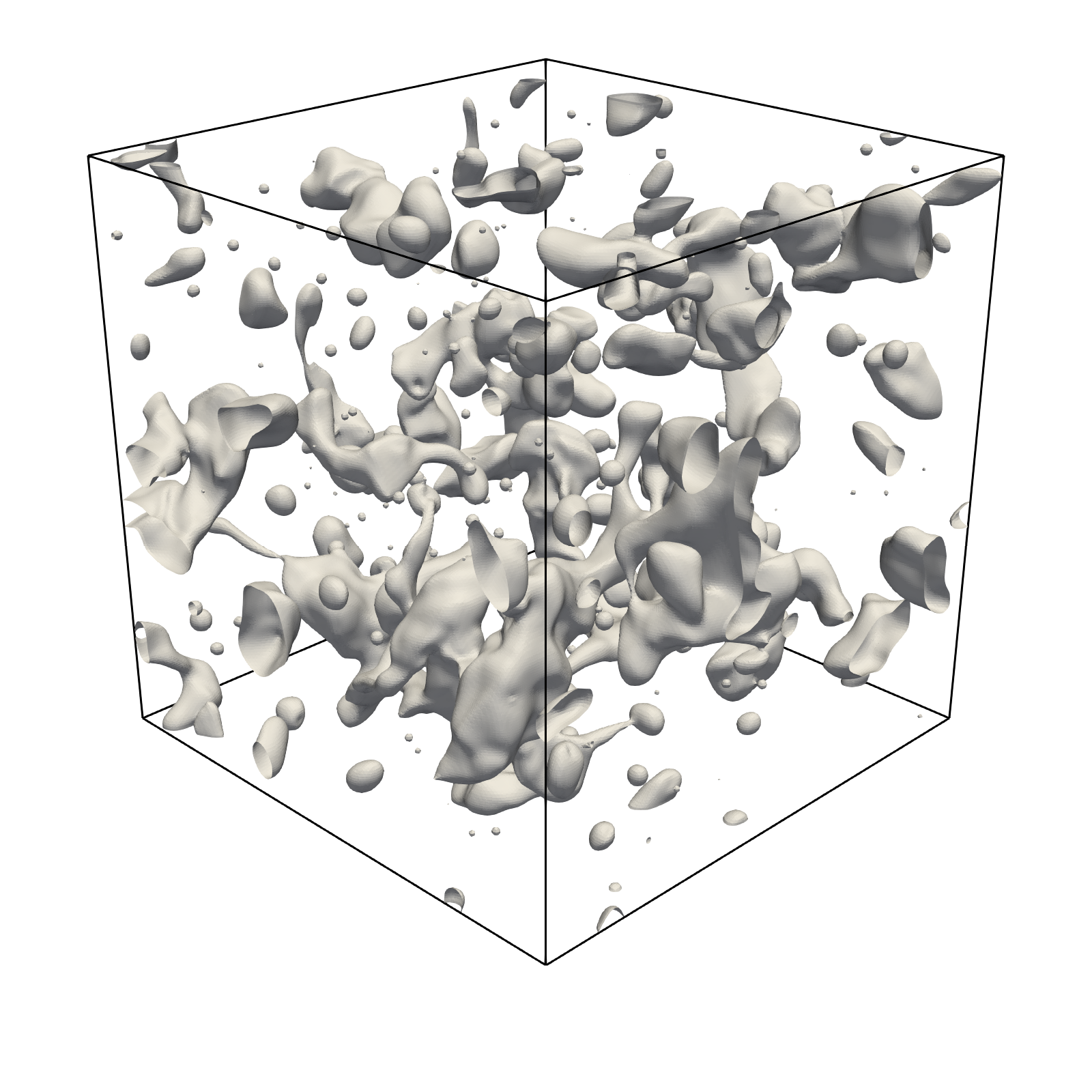}
        \caption{High $We_l$ regime}
        \label{fig:phi 9_5}
    \end{subfigure}

    \caption{\label{fig:phi_iso} Isocontours of constant volume fraction ($\phi = 0.5$) in the (a) low Weber number regime ($We_l = 2.0$) and (b) high Weber number regime ($We_l = 9.5$)}
\end{figure}

In the out-of-distribution case, the structures visible on the left boundary of Figs.~\ref{fig:pred}(a-c) are not isolated drops, but rather 2D planar projections of large corrugations from the primary droplet wrapping across the periodic boundary (corroborated by the 3D DNS volume fraction isocontours in Fig.~\ref{fig:phi 2_0}). The data-driven model struggles to predict the corrugations (red box, Fig.~\ref{fig:pred} (b)), diffusing them, and predicting non-physical interfaces. By contrast, the physics-based model (green box, Fig.~\ref{fig:pred}(c)) suppresses this diffusion better and shows an improvement in capturing the sharp features. 
Notably, both models fail to capture the smallest drops (in Fig.~\ref{fig:phi 2_0}) that pinched off the primary droplet at $We_l = 2.0$. These isolated droplets are physically smaller than the LES grid (downsampled) resolution. Consequently, their associated subgrid interfacial area lacks physical significance in the fractal sense, where part of the area is resolved in the LES grid. The inability to capture the smallest drops is also not critical to the predictive performance of the models, as they do not contribute significantly to the total interfacial area~\cite{hatashita2025scalings}. Modeling predictions of entirely subgrid bubbles/droplets would have to rely on different physics, such as total area scalings~\cite{jain:2025b}.

\begin{figure}
\centering
\includegraphics[width=\textwidth]{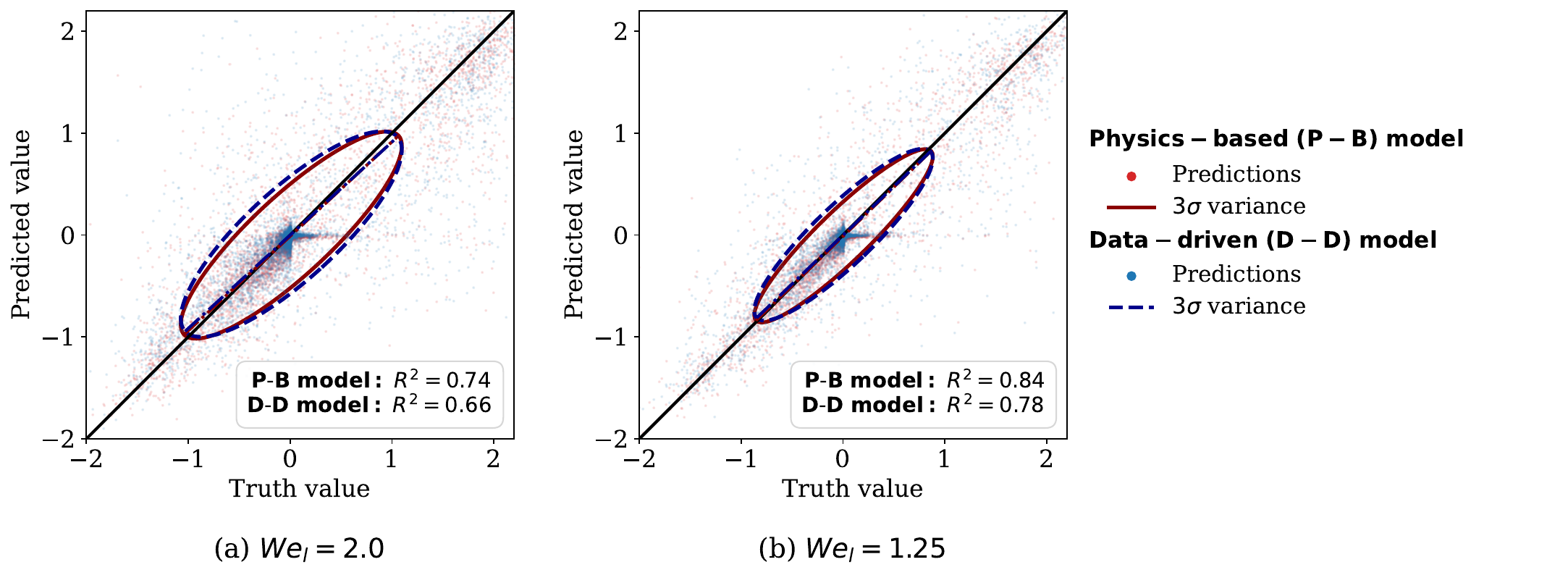}
\caption{\label{fig:R2} Coefficient of determination ($R^2$) between the prediction of subgrid interfacial area density from the data-driven model and the physics-based model compared against the ground truth for (a) $We_l = 2.0$ and (b) $We_l = 1.25$.}
\end{figure}

These qualitative observations are supported by $R^2$ scores and error variances  for both test sets as shown in Fig.~\ref{fig:R2}. These plots also include the $3\sigma$ (three standard deviations) covariance ellipses (which cover $99.7 \%$ of the data) to visually bound the predictive errors. For $We_l = 1.25$, the data-driven model achieves an $R^2$ score of $0.78$ and a prediction error variance of $0.0185$. The addition of the fractal-based loss term improves the predictions, by improving the $R^2$ to $0.84$, and reducing the variance to $0.0138$. The gap in generalization performance is also seen when extrapolating to $We_l = 2.0$. Here, the data-driven model has an $R^2$ of $0.66$ and an error variance of $0.0450$, while the physics-based model has an $R^2$ of $0.74$ and a variance of $0.0338$. The smaller statistical spread of the physics-based model predictions reflects the improved capturing of the sharp subgrid corrugations seen in the $2D$ planes (red and green boxes) in Fig.~\ref{fig:pred}.  

The distinct spread of predictions along the zero-axes (when either the predicted value or the truth value of the subgrid interfacial area density is approximately zero) offers further insight into these errors. Since neural network predictions are continuous and never exactly zero, even in the bulk fluid, we apply a tolerance value, $\epsilon=0.05$ to evaluate the predictions as a binary classification problem. Both predictions and ground truth below $\epsilon$ are classified as zero subgrid area (bulk fluid), while predictions above it are classified as true subgrid interfaces. We used different $\epsilon$ values and observed the same trend discussed below.

The vertical spread corresponds to false positives, where the true subgrid interfacial area density is zero, but the models predict non-zero values. At $We_l = 1.25$, the data-driven model has a false positive rate of $2.32 \%$, which increases to $3.92 \%$ at $We_l = 2.0$ reflecting the non-physical interface predictions previously observed. The physics-based model has lower false positive rates of $2.14 \%$ and $3.60 \%$, respectively, implying lower hallucinations.
The horizontal spread on the other hand represents false negatives, where the models fail to capture the subgrid interfacial area density present in the ground truth (specifically the smaller droplets discussed earlier). The physics-based model maintains lower false negative values for both in-distribution ($16.37 \%$ vs $17.54 \%$) and out-of-distribution sets ($15.70 \%$ vs $16.98 \%$), showing it is more successful at preserving the area of the resolved interfaces. 

\begin{figure}
\centering
\includegraphics[width=\textwidth]{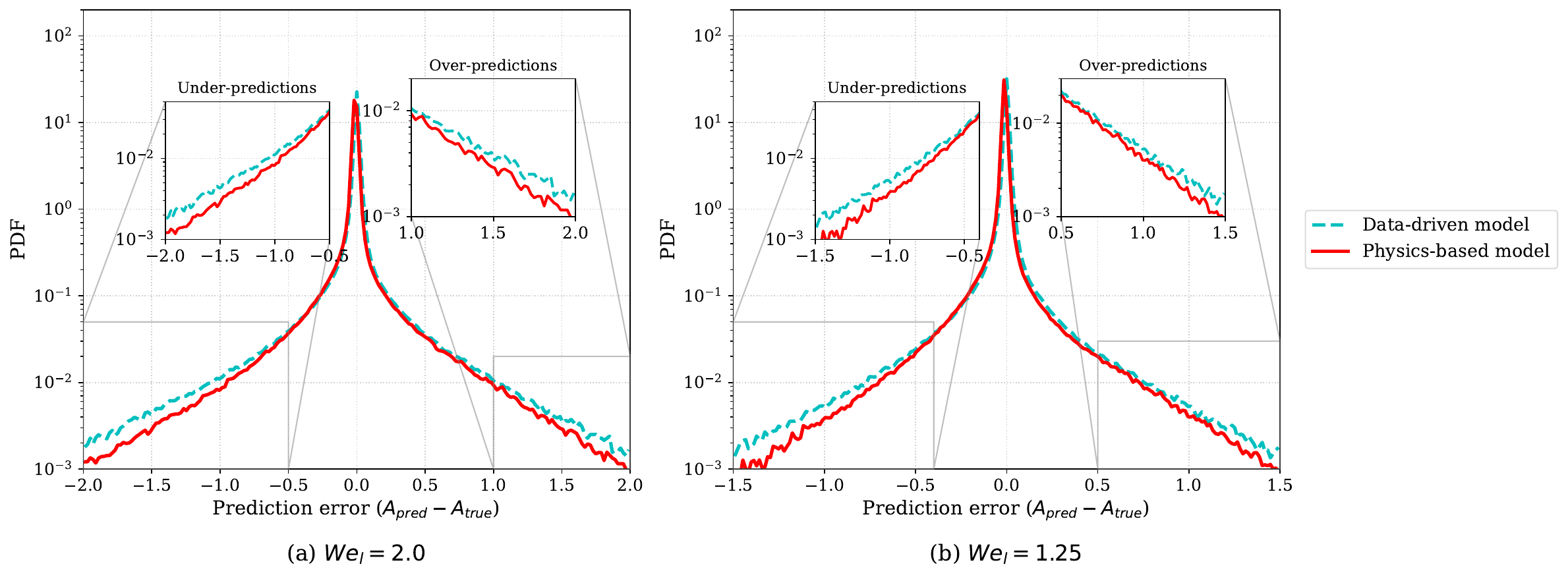}
\caption{\label{fig:PDF} Probability density function (PDF) of the prediction error for the subgrid interfacial area density for the data-driven model and the physics-based model at (a) $We_l = 2.0$ and (b) $We_l = 1.25$.}
\end{figure}

We also examine the probability density function (PDF) of the prediction errors for both Weber number test cases in Fig.~\ref{fig:PDF}. Both models show a peak near zero, which confirms that the vast majority of bulk fluids and interfaces are predicted correctly. While the purely data-driven model achieves a marginally higher probability density at the peak, indicating a slight over-optimization in the bulk fluid, the fundamental distinction between the models emerges in the error tails.
The right inset in Fig.~\ref{fig:PDF} focuses on significant over-predictions. This corresponds to the smearing of unphysical subgrid area into the surrounding bulk fluid as observed for the data-driven model in Fig.~\ref{fig:pred}. The left inset on the other hand are severe under-predictions which physically represent regions where the models fail to capture the corrugations on the surface.
Across both Weber numbers, the data-driven model shows a slightly higher probability for such occurrences. The physics-based model outperforms it by capturing more corrugations (left tail) and suppressing unphysical interfacial smearing (right tail). As expected, the tails are wider for $We_l = 2.0$, since the reduced surface tension at this Weber number causes more wrinkling. Ultimately, since the $R^2$ scores are affected by large squared errors, the data-driven model's marginal superiority in the trivial bulk fluid is overshadowed by its higher error at the interfaces.

\begin{figure}
    \centering
    \begin{subfigure}[b]{0.49\textwidth}
        \centering
        \includegraphics[width=\textwidth]{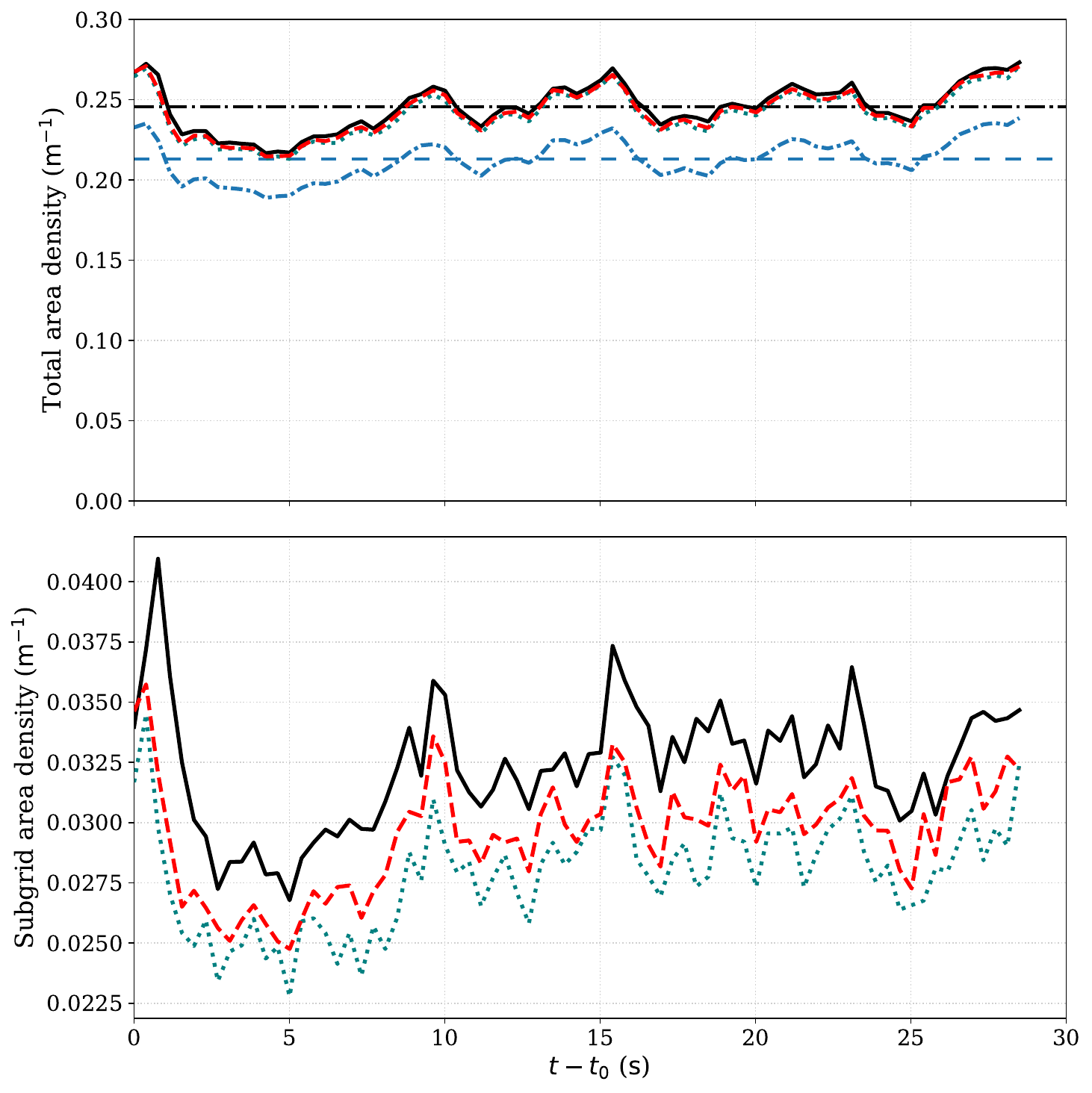}
        \caption{$We_l = 2.0$}
        \label{fig:curve-2_0}
    \end{subfigure}
    \hfill
    \begin{subfigure}[b]{0.49\textwidth}
        \centering
        \includegraphics[width=\textwidth]{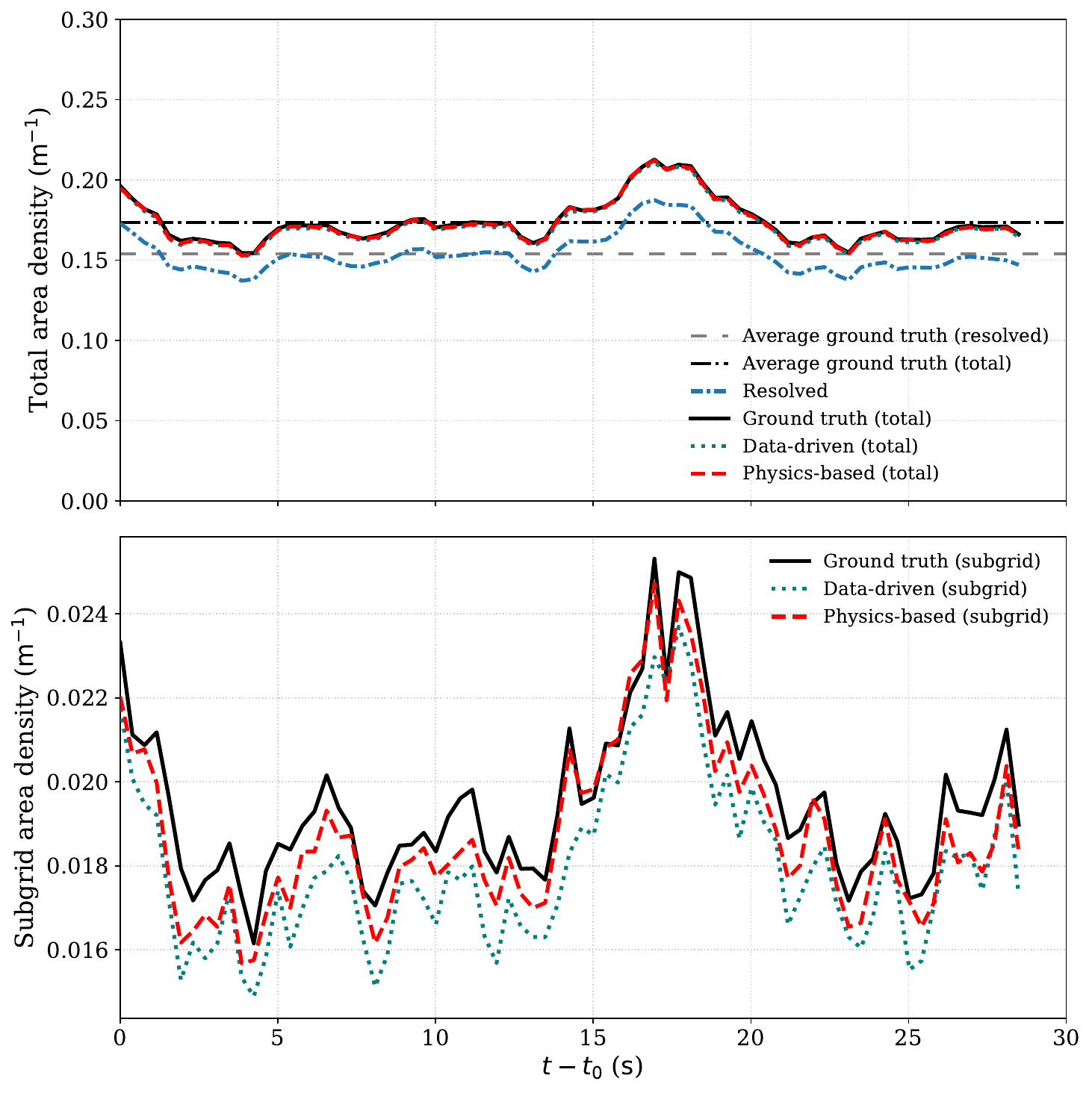}
        \caption{$We_l = 1.25$}
        \label{fig:curve-1_25}
    \end{subfigure}
    \caption{\label{fig:area} Comparison of the global total and subgrid interfacial area densities estimated from the data-driven model and the physics-based model against the ground truth for (a) $We_l = 2.0$, and (b) $We_l = 1.25$, as a function of time.}
\end{figure}

The global total interfacial area density (total area divided by the domain volume), which is the sum of the resolved and the subgrid area densities, is an important metric for modeling. We restrict our integration to the droplet interfaces, and exclude the resolved bulk fluid (with zero subgrid area).
For a well-resolved large-eddy simulation at the low Weber number regime with a downsampling factor of $4$, it is expected that the resolved area density captures most of the interfacial area density. Figure~\ref{fig:area} shows how the global total interfacial area density varies with time for both test sets ($t_0$ indicates the time associated with the first snapshot in the test set). Since the flow is stationary, the area is expected to oscillate about the ensemble average area, which is our quantity of interest. The elapsed time in the test dataset is approximately $20$ eddy turnover times. For $We_l = 1.25$, the resolved area density is $0.154$ m$^{-1}$, whereas the subgrid area density is $0.02$ m$^{-1}$, yielding a true total interfacial area density of $0.174$ m$^{-1}$. The predictive closure models should then ideally capture $11.5 \%$ of the total area density. The purely data-driven model predicts a subgrid area density of $0.018$ m$^{-1}$, yielding a total area density of $0.172$ m$^{-1}$ (recovering $98.8 \%$ of the ground truth), while the physics-based model improves on this by predicting $0.019$ m$^{-1}$, and a total area density of $0.173$ m$^{-1}$, achieving a $99.4 \%$ recovery. At the higher $We_l =2.0$, the resolved and the subgrid area densities increase to $0.214$ m$^{-1}$ and $0.032$ m$^{-1}$ respectively due to weaker surface tension forces, resulting in a global total area density of $0.246$ m$^{-1}$ across the dataset (with the subgrid contribution rising to $13 \%$). The data-driven model predicts a subgrid area of $0.028$ m$^{-1}$, yielding a total area of $0.242$ m$^{-1}$ (recovering $98.4 \%$ of the ground truth), while the physics-based model predicts a subgrid area of $0.03$ m$^{-1}$, achieving a total area of $0.244$ m$^{-1}$(recovering $99.2 \%$ of the ground truth). All the reported values are averaged across the entire test dataset (ensemble average). This shows that, by soft constraining the model to respect the underlying fractal dimension of the interface, the physics-based model outperforms its purely data-driven counterpart.

\subsection{Comparison with standard regularizer}

To check if our fractal term in Eq.~\eqref{eq:physics_loss} is merely acting as a mathematical regularizer, we compared both models against a data-driven model trained with a standard AdamW~\cite{loshchilov:2019} optimizer, which includes decoupled weight decay ($L_2$ regularization). While the AdamW model improves $R^2$ to $0.72$ for $We_l = 2.0$ and $0.82$ for $We_l = 1.25$, compared to the model with no regularizer ($0.66$ and $0.78$, respectively) by shrinking model weights and reducing overall variance, the physics-based model still outperforms in $R^2$ scores ($0.74$ and $0.84$, respectively). 

\begin{figure}
    \centering
    \begin{subfigure}[b]{0.32\textwidth}
        \centering
        \includegraphics[width=\textwidth]{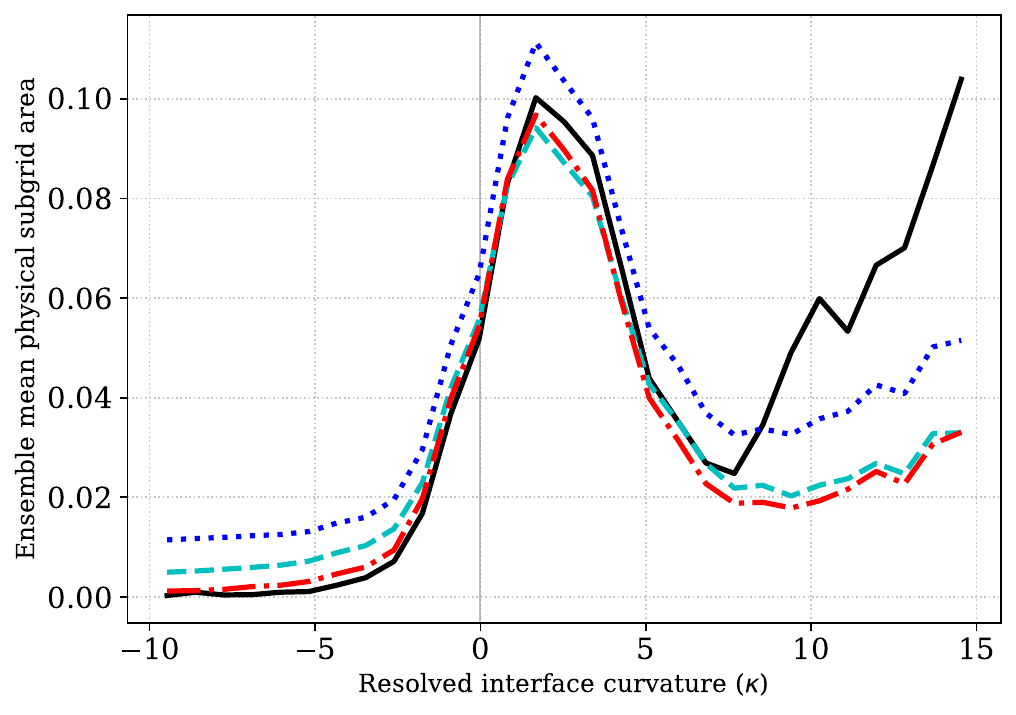}
        \caption{$We_l = 2.0$}
        \label{fig:curve-2_0}
    \end{subfigure}
    \hfill
    \begin{subfigure}[b]{0.32\textwidth}
        \centering
        \includegraphics[width=\textwidth]{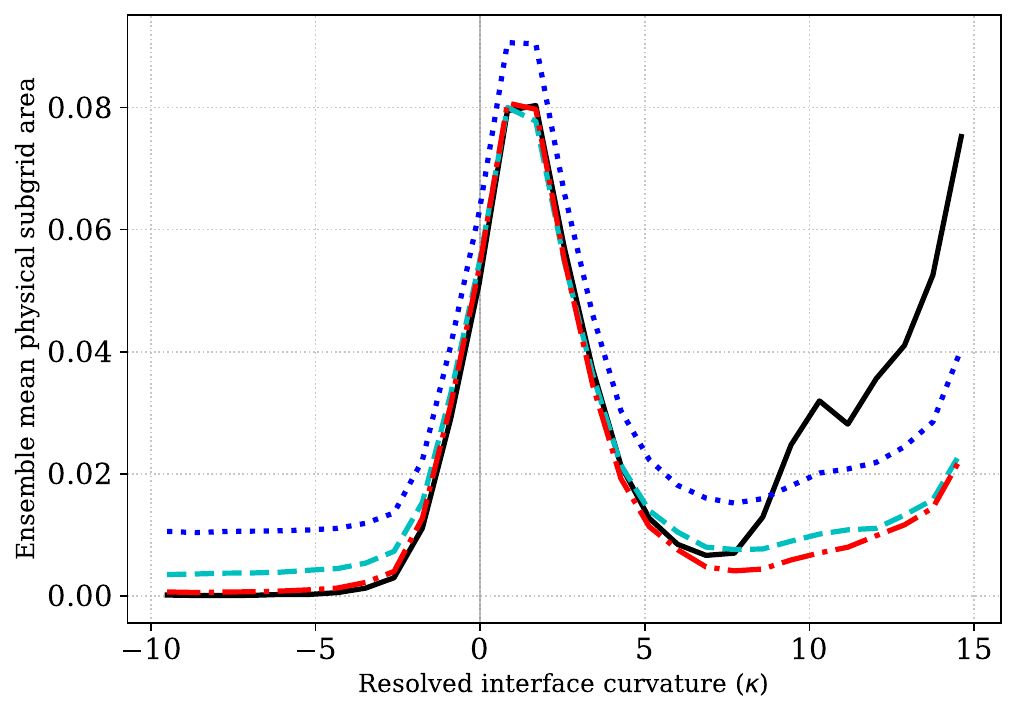}
        \caption{$We_l = 1.25$}
        \label{fig:curve-1_25}
    \end{subfigure}
    \begin{subfigure}[b]{0.32\textwidth}
        \centering
        \includegraphics[width=\textwidth]{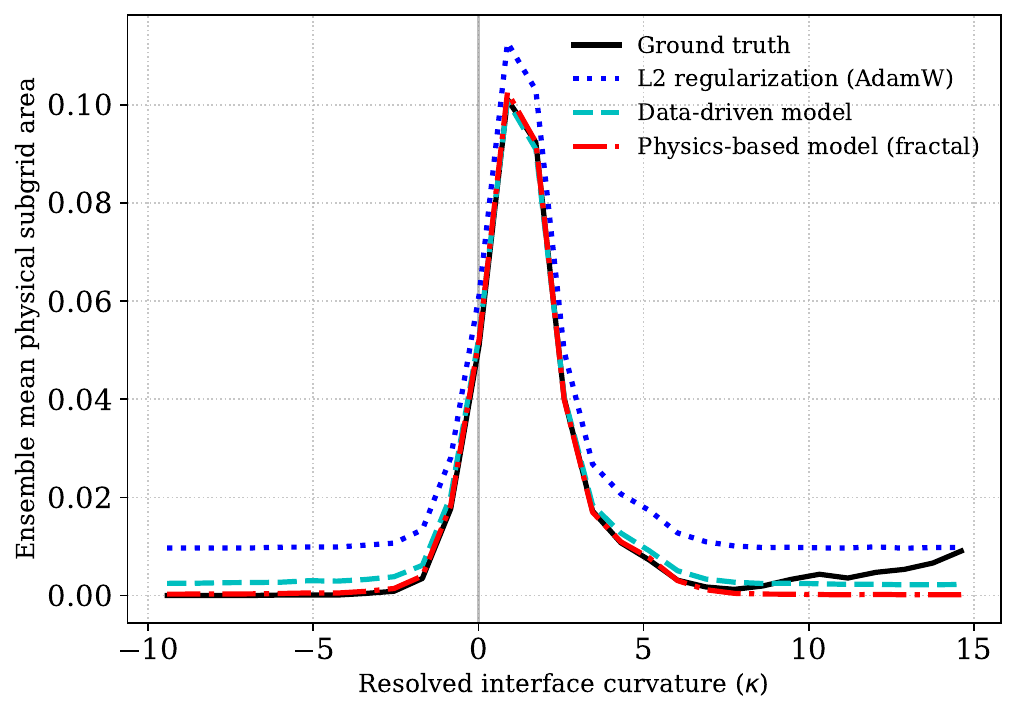}
        \caption{$We_l = 0.50$}
        \label{fig:curve-0_5}
    \end{subfigure}

    \caption{\label{fig:curve} Comparison of the physical subgrid area density conditioned on the curvature predicted from the data-driven model and the physics-based model compared against the ground truth for (a) $We_l = 2.0$, (b) $We_l = 1.25$ and (c) $We_l = 0.50$.}
\end{figure}

The physics-based model really shines however when we calculate the ensemble mean physical subgrid area across the entire set as shown in Fig.~\ref{fig:curve}. In this context, the ensemble mean physical subgrid area denotes the expected value of the subgrid area conditioned on the local resolved curvature, $\hat \kappa = -\overline\nabla\cdot (\overline\nabla \;\overline\phi/|\overline\nabla\; \overline\phi|)$. It is calculated by averaging the conditioned subgrid area density across the entire ensemble of temporal snapshots, strictly bounding the predicted areas to non-negative values to keep the analysis focused on physical corrugations. $\hat \kappa \leq 0$ represents flat or slightly concave surfaces. This is where the AdamW regularization fails and over-predicts subgrid area density. Since standard $L_2$ regularization penalized large weights, the model ends up diffusing predictions and hallucinating unphysical area in these regions. The purely data-driven model does slightly better, but it still struggles to force the area down to zero for negative values of $\hat\kappa$ ($<5$), meaning it still predicts unphysical areas. The physics-based model successfully suppresses this error, tightly tracking the ground truth.
$0<\hat\kappa<5$ corresponds to the regions with the maximum corrugations, which has the maximum contribution to the subgrid area density. The AdamW model over-predicts the peak, whereas both the purely data-driven and physics based models capture it quite accurately.
Finally, at higher curvatures ($\hat\kappa > 10$), all the models diverge from the ground truth. This is represented significantly by small droplets and capillary threads that are physically smaller than the coarse LES grid. To confirm that this divergence is driven by fragmentation, we also tested our models on $We_l = 0.5$ in Fig.~\ref{fig:curve}(c). The surface tension forces here are much stronger and they actively resist droplet breakup. This is marked by the absence of the large area tail at higher curvatures in the ground truth. The resolved quantities as inputs do not have enough information to capture subgrid area density at that exact location for $We_l = 1.25$ and $2.0$. The fractal model captures the dominant structures that make up the vast majority of the droplet area, making it the best model out of the three for the closure problem that we are studying. 

\subsection{High $We_l$ regime}
We carried out a similar analysis at the high Weber number regime where the inertia forces dominate surface tension forces which leads to significant interface fragmentation. The datasets of $900$ snapshots in this regime are filtered and downsampled from the DNS simulations at $Re_\lambda= 87$ by a factor of $8$ from $256^3$ grid points to $32^3$ grid points. For this analysis, the models were trained on an interpolation dataset featuring high Weber numbers $(We_l \in \{4.0,6.5,14.0\})$ and evaluated on an unseen test set at $We_l = 9.5$.

While the lower $We_l$ regime is dominated by corrugations on the surface with rare breakup events due to intermittency, the interface in the high $We_l$ regime undergoes significant breakup, generating many spherical sub-Kolmogorov-Hinze interfaces. Since the physics-based model is dictated by a fractal dimension $(>2)$ that describes continuous corrugated surfaces, its physical assumptions weaken when applied to spherical droplets (where the fractal dimension is $\approx2$).

\begin{figure}
\centering
\includegraphics[width=0.95\textwidth]{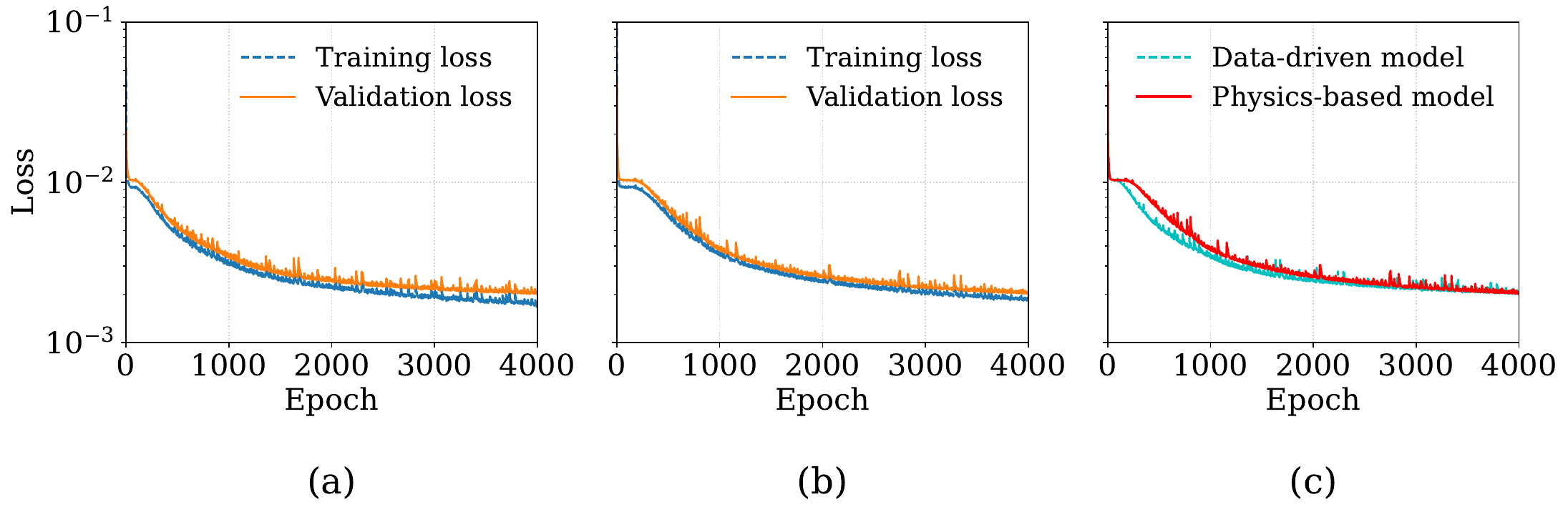}
\caption{\label{fig:Loss high We} Training and validation loss versus the number of epochs in the high $We_l$ regime for (a) Data-Driven Model and (b) Physics-Based Model. (c) The physics-based and data-driven models have similar validation losses.}
\end{figure}

\begin{figure}
\centering
\includegraphics[width=\textwidth]{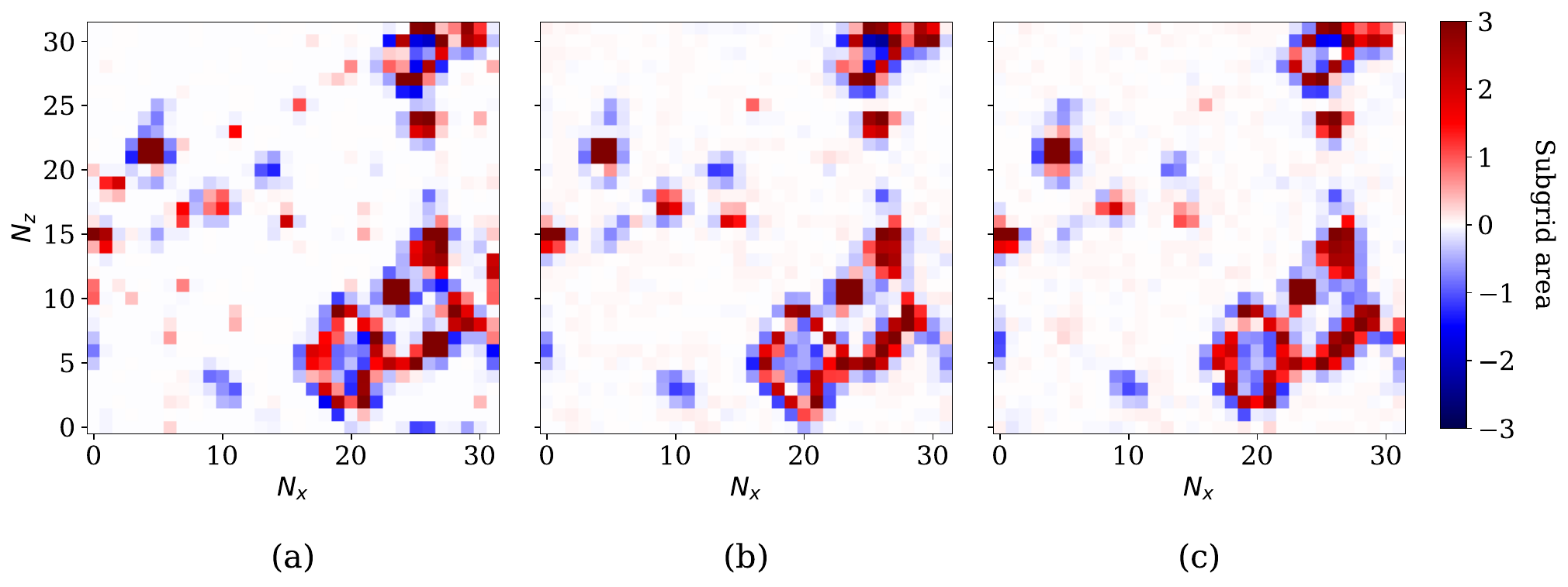}
\caption{\label{fig:pred high We}Qualitative comparison between the prediction of subgrid interfacial area density (a) from the ground truth, (b) the data-driven model, and the (c) physics-based model at the high $We_l$ regime $(We_l = 9.5)$. The central $x-z$  plane is shown in each case.}
\end{figure}

This is reflected in Fig.~\ref{fig:Loss high We} where both the data-driven and physics-based models show nearly identical training and validation losses. The validation losses for both models in Fig.~\ref{fig:Loss high We} are indistinguishable. Furthermore, the data-driven and physics based models have similar $R^2$ scores of $0.80$ and $0.81$, respectively. 
This is further corroborated by the qualitative observations in Fig.~\ref{fig:pred high We}. Unlike the low $We_l$ regime where the physics-based model clearly captured corrugations better, the predictions in this regime show no discernible difference. Both models capture the primary structures but struggle when attempting to predict the highly fragmented smaller droplets.

This also shows that the fractal-regularizer in the low $We_l$ regime is not merely a general statistical regularizer, otherwise it would have improved performance regardless of the regime. It improves predictions when the surface tension forces are stronger maintaining continuous corrugated surfaces where the fractal theory is valid, and it becomes inert when the flow physics transitions into chaotic atomization and breakup with weaker surface tension forces.

\section{Conclusion}
\label{sec:conclusions}
In this work, we developed and evaluated first-of-its-kind machine learning models for predicting the local and global subgrid interfacial area density in three-dimensional two-phase turbulent flows. We introduced a purely data-driven model based on an encoder–decoder architecture with skip connections, alongside a physics-based model incorporating a regularization term derived from fractal interface theory.

In the low Weber number regime, both models recover the global total interfacial area density for both in-distribution and out-of-distribution datasets ($> 98\%$ for the data-driven model and $> 99\%$ for the physics-based model). The physics-based model achieves consistently higher $R^2$ values ($> 8\%$ improvement over its purely data-driven counterpart), while also reducing error variance, suppressing nonphysical local subgrid area density, and capturing sharper interfacial corrugations. Furthermore, it outperforms standard regularization approaches (\textit{e.g.}, AdamW) in predicting physically consistent subgrid area density conditioned on resolved interface curvature, particularly in flat and concave regions. These improvements arise from embedding the fractal scaling behavior of corrugated interfaces into the learning process.

In contrast, in the high Weber number regime dominated by strong fragmentation into sub-Kolmogorov-Hinze interfaces, the physics-based model does not show improvement over the data-driven model. This demonstrates that the fractal-based constraint does not merely act as a generic regularizer, but rather derives its effectiveness from the validity of its underlying physical assumptions. When the interface morphology transitions from corrugated structures to nearly spherical fragments, the imposed inductive bias becomes less relevant, and both models exhibit similar performance.

Beyond this specific application, this work highlights a broader paradigm for scientific machine learning: the effectiveness of physics-informed models depends critically on the alignment between embedded inductive biases and the governing physical regime. Our results suggest that future subgrid closures (and, more generally, AI models for multiscale systems) should incorporate regime-aware physical constraints rather than static assumptions. Such approaches offer a pathway toward more robust, generalizable, and physically consistent learning frameworks for complex turbulent and multiphase flows. These findings suggest that the future of scientific machine learning lies in designing models whose inductive biases evolve with the physics they seek to represent.

\section*{Data and Code Availability}
The machine learning training scripts and model architectures developed for this study will be made publicly available in a persistent repository upon acceptance of the manuscript. 

\begin{acknowledgments}
The authors acknowledge the support from the George W. Woodruff School of Mechanical Engineering at Georgia Institute of Technology, and a partial support from the GTRI-GWW Connect grant.
S.~S.~J. acknowledges support by the donors of ACS Petroleum Research Fund under Doctoral New Investigator Grant 69196-DNI9 (S.~S.~J. served as Principal Investigator on ACS PRF 69196-DNI9).

The authors also acknowledge the generous computing resources from the DOE's 2024 and 2025 ALCC awards (TUR147 \& BubbleLaden, PI: Jain). This research used supporting resources at the Argonne and the Oak Ridge Leadership Computing Facilities. The Argonne Leadership Computing Facility at Argonne National Laboratory is supported by the Office of Science of the U.S. DOE under Contract No. DE-AC02-06CH11357. The Oak Ridge Leadership Computing Facility at the Oak Ridge National Laboratory is supported by the Office of Science of the U.S. DOE under Contract No. DE-AC05-00OR22725.
Finally, the authors acknowledge the contribution of Roy Mazor, who is an undergraduate researcher in the Flow Physics and Computational Science Lab for carrying out the filtering analysis alongside the authors. Preliminary results for this work were presented at the AIAA SciTech 2026 Conference~\cite{bhattacharjee:2026}.
\end{acknowledgments}

\appendix

\section{On filtering diffuse interfaces}
\label{apx:filtering}

Local interfacial area density is defined as
\begin{equation}
    \label{eq:def-ia} 
    \delta = |\nabla \phi|,
\end{equation}
where $\phi$ is volume fraction field, used as the indicator function. This equation follows from the co-area formula, which relates the total variation of a function of bounded variation [$\|\nabla \phi\| (\Omega)$, where $\phi \in L^1_{\rm loc} (\Omega), BV(\Omega)$] to the perimeter of a set [$P(\Omega;\phi)$]~\cite{leoni:2009}. Here, total variation of $\phi$ over the domain is taken to be $\int_{L^3} |\nabla \phi | dV$, which is equal to the total area of the interface by the co-area formula. We take $\delta$ to be the local area contribution over a control volume.

\textit{A-priori} analysis is performed on the filtered volume fraction field, \textit{i.e.}, filtered DNS data is compared against filtered LES field. The filtered DNS interfacial area is computed based on the physical definition of interfacial area density, namely that area density is the ratio of total area within a unit volume. The total interfacial area follows from the integral of $\delta$ in a control volume. Thus, the filtered DNS interfacial area is obtained by
\begin{equation}
    \label{eq:delta_dns_filt}
    \langle {\delta_{\rm DNS}}\rangle = \frac{1}{\overline \Delta_{\rm DNS}^3} \int_{\overline \Delta_{\rm DNS}^3} \delta dV = \frac{1}{\overline \Delta_{\rm DNS}^3} \int_{\overline \Delta_{\rm DNS}^3} \left | \nabla \phi \right | dV,
\end{equation}
where $\overline \Delta$ is the filter width (here $\overline{\Delta}_{\rm DNS} = \overline{\Delta}$). This expression guarantees conservation of the total interfacial area in the domain. 

LES is performed on coarser grids, therefore, the volume fraction is first filtered and coarsened, $\overline \phi$, and the correspondent interfacial area density is computed from the gradient of the filtered volume fraction [Eq.~\eqref{eq:def-ia}], 
\begin{equation}
    \label{eq:delta_les} 
    \delta_{\rm LES} = \left | \overline \nabla \; \overline \phi \right |,
\end{equation}
where the filtering operation on the volume fraction, is based on a convolution kernel $G(r;\overline\Delta)$, defined in Eq.~\eqref{eq:filt-def}, and the coarsening and the gradient operators ($\overline{\nabla}$) are defined on the LES grid.

\begin{figure}
    \centering
    \begin{tikzpicture}
        \matrix[column sep=2pt] (m) {
            \node[inner sep=0] (A0) {\includegraphics[width=0.47\textwidth]{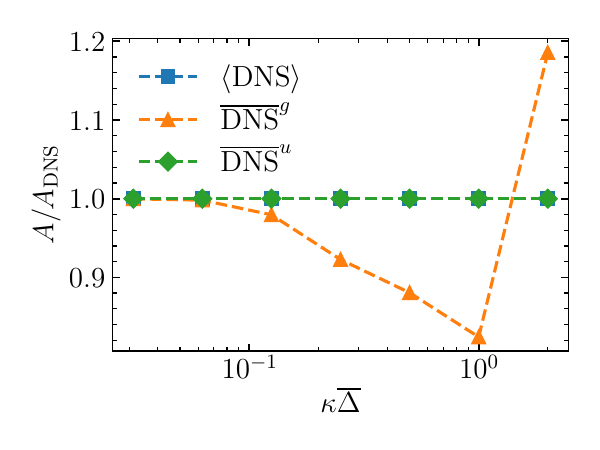}}; &
            \node[inner sep=0] (A1) {\includegraphics[width=0.47\textwidth]{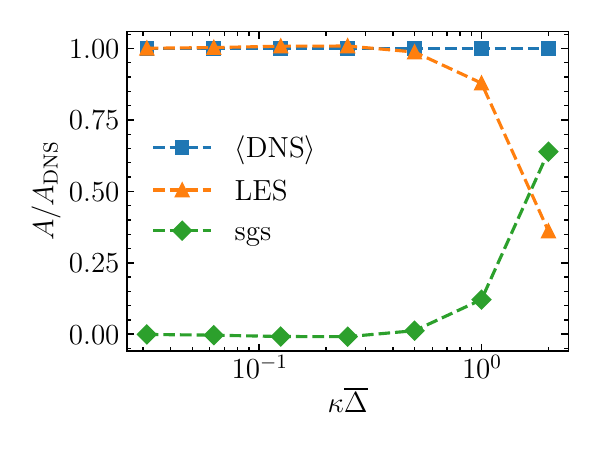}}; \\
        };

        \begin{scope}[shift={(m.south west)}, x={(m.south east)}, y={(m.north west)}]
            
            \node[black, font=\boldmath, anchor=south] at (0.05, -0.05) {(a)};
            
            \node[black, font=\boldmath, anchor=south] at (0.55, -0.05) {(b)};

        \end{scope}
    \end{tikzpicture} 
    \caption{Comparison of filtering procedures for a non-zero curvature surface. (a) Evaluation of DNS filtering approaches. (b) Evaluation of total resolved and subgrid interface area compared against DNS.}
    \label{fig:cons-area-proc-circ2d}
\end{figure}
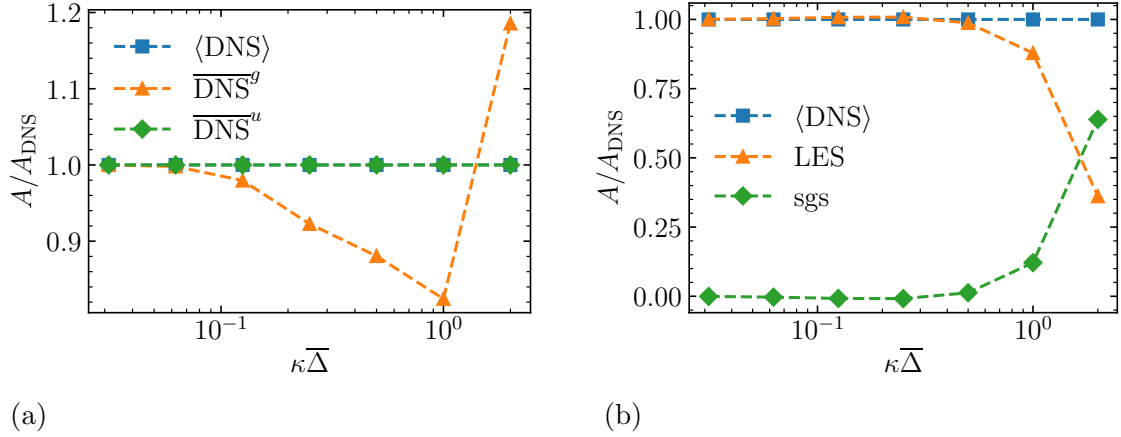

First, we may demonstrate that inappropriate filtering of DNS yields in the loss of total interfacial area. Take a non-zero curvature circular interface in 2D defined by
\begin{equation}
\label{eq:circ2d-prof}
\phi = \frac{1}{2} \left [1 + \tanh \left (\frac{r-R}{2\epsilon} \right ) \right ],
\end{equation}
where $r = \sqrt{x^2+y^2}$ is the local radius, $R$ is the circle radius, and $\epsilon$ is the interface thickness, taken to be equal to the grid size. Figure~\ref{fig:cons-area-proc-circ2d}(a) demonstrates the effects of standard filtering compared to the proposed approach. For both the uniform top-hat filter kernel ($\overline{\rm DNS}^u$) followed by coarsening and the volume averaged approach ($\langle {\rm DNS} \rangle$), total interfacial area is conserved across a wide range of filter widths ($\overline{\Delta}_{\rm DNS}$) if compared to the DNS interfacial area, this is due to the fact that for uniform grid spacings, the top-hat filter kernel reduces to a volume averaging operation. Nevertheless, the use of a Gaussian filter followed by a coarsening operation ($\overline{\rm DNS}^g$) mispredicts total interfacial area.
Second, we demonstrate that actual total subgrid interfacial is only generated for sufficiently large filter widths. Figure~\ref{fig:cons-area-proc-circ2d}(b) compares the volume averaged DNS total interfacial area ($\langle {\rm DNS} \rangle$) with the resolved LES area (LES, where we take $\overline{\Delta}_{\rm DNS} = \overline{\Delta}_{\rm LES} = \overline{\Delta}$), and their difference is defined as the subgrid scale total interfacial area contribution (sgs). We observe that, only for $\kappa \overline{\Delta} \gtrapprox 1$, there is a significant deficit in total interfacial area. For the proposed interface profile [Eq.~\eqref{eq:circ2d-prof}], where the radius of the surface is constant, the interface curvature $\kappa$ is also a constant, $\kappa = 1/R$. Therefore, the loss of interfacial area only occurs when the filter width is comparable to the radius of the surface. For the constant radius circle, the loss of interfacial area occurs when the entire circle is subgrid. Nevertheless, in turbulence, due to the wide range of scales, it is expected that larger-magnitude curvatures (meaning either very small spherical interfaces of small scale corrugations) will be lost earlier for smaller filter widths, whilst part of the large interface is resolved on the LES grid.

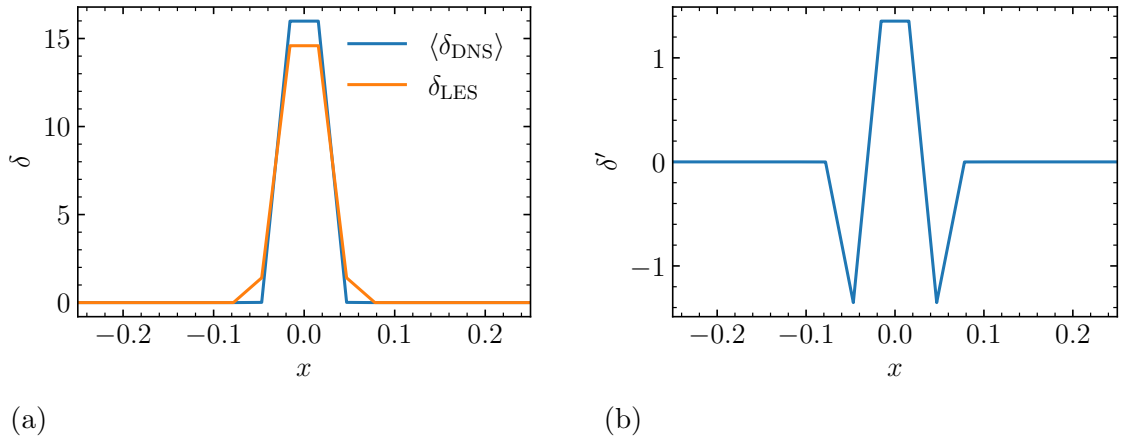
\begin{figure}
    \centering
    \begin{tikzpicture}
        \matrix[column sep=2pt] (m) {
            \node[inner sep=0] (A0) {\includegraphics[width=0.47\textwidth]{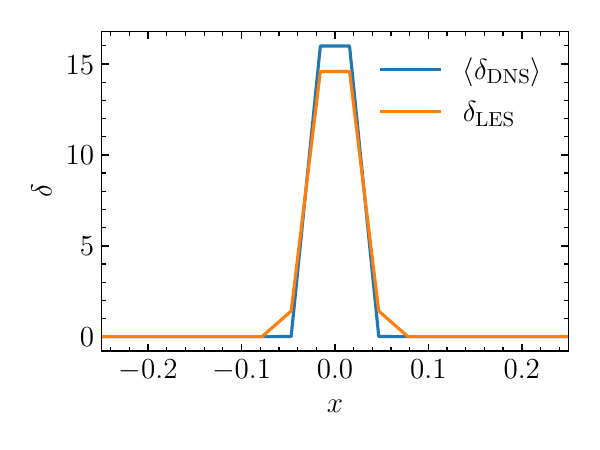}}; &
            \node[inner sep=0] (A1) {\includegraphics[width=0.47\textwidth]{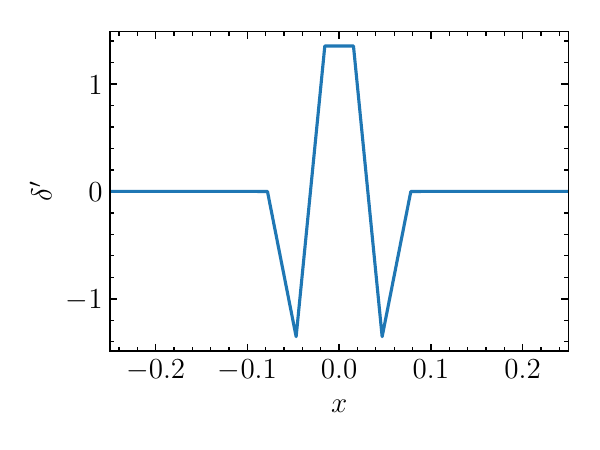}}; \\
        };

        \begin{scope}[shift={(m.south west)}, x={(m.south east)}, y={(m.north west)}]
            
            \node[black, font=\boldmath, anchor=south] at (0.05, -0.05) {(a)};
            
            \node[black, font=\boldmath, anchor=south] at (0.55, -0.05) {(b)};

        \end{scope}
    \end{tikzpicture} 
    \caption{(a) Comparison of filtered interfacial area, and (b) subgrid interfacial area for standard filtering procedure, where $\overline \Delta_{\rm DNS} = \overline \Delta_{\rm LES}$.}
    \label{fig:issues-filt-interface}
\end{figure}

\begin{figure}
    \centering
    \includegraphics[width=0.5\linewidth]{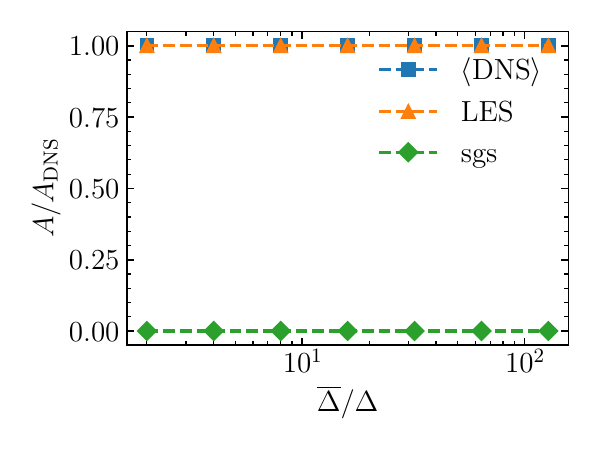}
    \caption{Comparison of total resolved and subgrid interfacial area against DNS for a zero-curvature surface.}
    \label{fig:cons-area-proc-1d2d}
\end{figure}

We have observed the generation of unphysical subgrid interfacial area density ($\delta'<0$)~\cite{hatashita:2025}, using the aforementioned procedure. And we have further identified the source of this problem, it arises from the discrepancy in interface thickness ($\overline{\epsilon}$) scales from the volume averaged DNS and the resolved LES interfacial area density. For instance, take a 1D interface, defined by
\begin{equation}
\label{eq:1d1d-prof}
\phi = \frac{1}{2} \left [1 + \tanh \left (\frac{x}{2\epsilon} \right ) \right ],
\end{equation}
which is representative of a zero-curvature surface. We demonstrate this effect in the Fig.~\ref{fig:issues-filt-interface}, where $\overline{\epsilon}_{\rm DNS} < \overline{\epsilon}_{\rm LES}$, therefore yielding in $\delta'<0$.
It is noteworthy that although there exists regions with $\delta'<0$, if we evaluate the total interfacial area to be capture with what is captured by the LES grid, that there is no net subgrid interfacial area, as observed in the Fig.~\ref{fig:cons-area-proc-1d2d}.

\begin{figure}
    \centering
    \begin{tikzpicture}
        \matrix[column sep=2pt] (m) {
            \node[inner sep=0] (A0) {\includegraphics[width=0.47\textwidth]{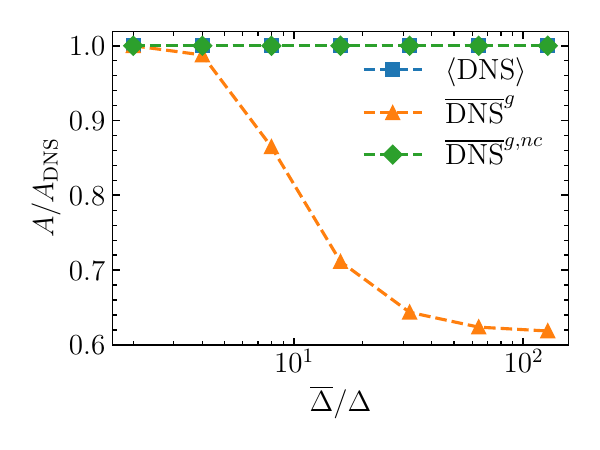}}; &
            \node[inner sep=0] (A1) {\includegraphics[width=0.47\textwidth]{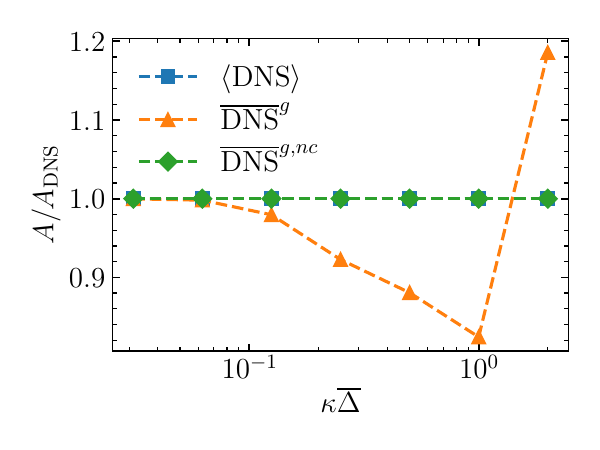}}; \\
        };

        \begin{scope}[shift={(m.south west)}, x={(m.south east)}, y={(m.north west)}]
            
            \node[black, font=\boldmath, anchor=south] at (0.05, -0.05) {(a)};
            
            \node[black, font=\boldmath, anchor=south] at (0.55, -0.05) {(b)};

        \end{scope}
    \end{tikzpicture} 
    \caption{Effect of coarsening on the loss of interfacial area using a Gaussian filter kernel for (a) zero-curvature, and (b) non-zero curvature surfaces.}
    \label{fig:effect-coarse}
\end{figure}
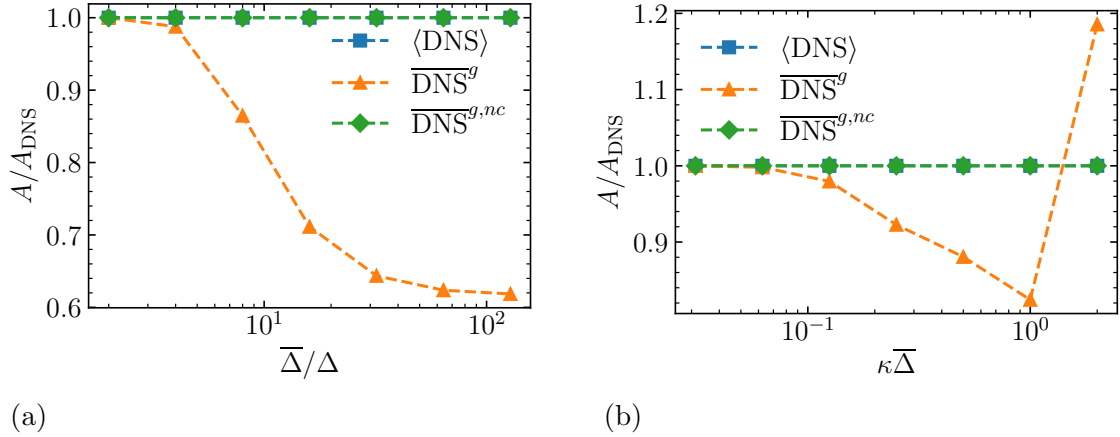

\begin{figure}
    \centering
    \begin{tikzpicture}
        \matrix[column sep=2pt] (m) {
            \node[inner sep=0] (A0) {\includegraphics[width=0.47\textwidth]{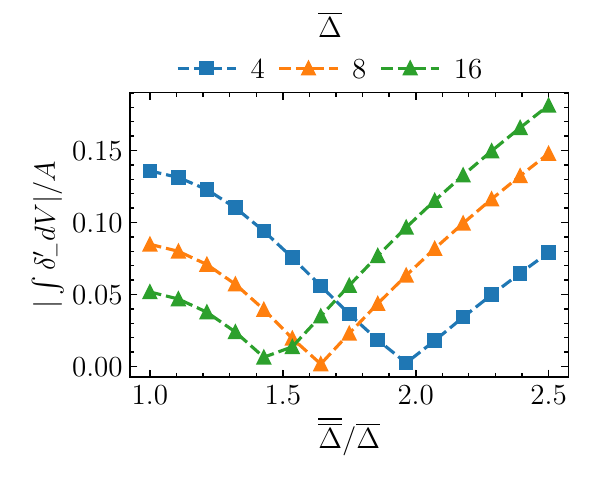}}; &
            \node[inner sep=0] (A1) {\includegraphics[width=0.47\textwidth]{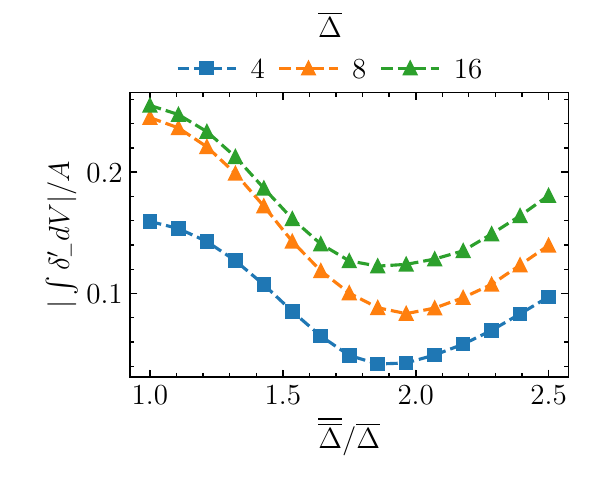}}; \\
        };

        \begin{scope}[shift={(m.south west)}, x={(m.south east)}, y={(m.north west)}]
            
            \node[black, font=\boldmath, anchor=south] at (0.05, -0.05) {(a)};
            
            \node[black, font=\boldmath, anchor=south] at (0.55, -0.05) {(b)};

        \end{scope}
    \end{tikzpicture} 
    \caption{Effect of the second filter width on the minimization of negative subgrid interfacial area for different underlying LES grids using a Gaussian filter kernel for (a) a zero-curvature, and (b) a non-zero curvature surfaces.}
    \label{fig:effect-second-filt}
\end{figure}
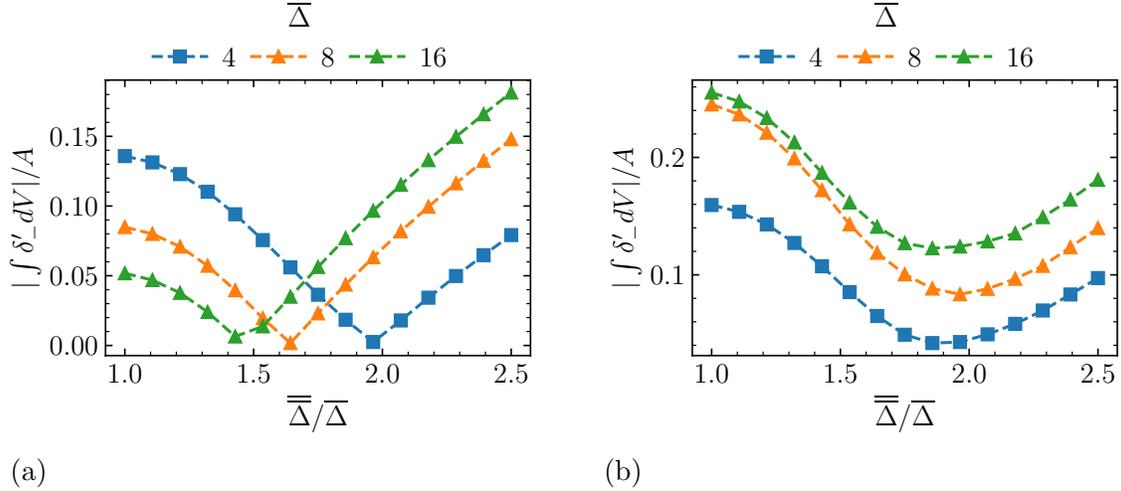

Some of the approaches to circumvent this issue is to either readjust the filtered interface thicknesses or to mask the model to the non-negative regions as proposed in the Section~\ref{subsec:models}. For instance, one approach to readjust the interface thicknesses is to diffuse the filtered DNS interface without losing interfacial area. We have tested passing $\langle \delta \rangle$ through a Gaussian filter without the coarsening step, this operation is defined as
\begin{equation}
    \label{eq:delta_dns_filt_filt} 
    \overline{\langle {\delta} \rangle } = \int_{-\infty}^{\infty} \mathbf{G}(\mathbf{r}, \overline{\overline{\Delta}}) \langle {\delta} \rangle (\mathbf{x}-\mathbf{r}) d\mathbf{r},
\end{equation}
where the second filter $\overline{\overline{\Delta}}$ has to be specified. Without coarsening ($\overline{\rm DNS}^{g,nc}$) for both zero and non-zero curvatures, in the Fig.~\ref{fig:effect-coarse}, we can observe no loss of interfacial area.
Nevertheless, upon verifying the required second filter width $\overline{\overline{\Delta}}$ to readjust the filtered interface thickness and therefore minimize negative subgrid interfacial area, we observed a variation of $\overline{\overline{\Delta}}$ for different cases. Figure~\ref{fig:effect-second-filt} demonstrates that, for both different cases and different underlying LES grid sizes, the second filter width, which minimizes the total negative subgrid interfacial area, changes. Therefore, we chose to adopt the masking procedure herein.

\bibliography{main}

\end{document}